\documentclass{article}

\usepackage[english]{babel}
\usepackage[letterpaper,top=2cm,bottom=2cm,left=3cm,right=3cm,marginparwidth=1.75cm]{geometry}
\usepackage{mathptmx}
\usepackage{sectsty}
\usepackage{amsmath}
\usepackage{bm}
\usepackage{graphicx} 
\usepackage{hyperref}

\usepackage[format=plain,justification=justified,singlelinecheck=false,font={stretch=1.125,small,sf},labelfont=bf,labelsep=space]{caption}

\usepackage{gensymb}
\usepackage{amsmath,amssymb}

\title{
Dynamical modes of highly elastic loops settling under gravity in a viscous fluid
}

\author{
  Yevgen Melikhov\textit{$^{a}$}
  and 
  Maria L. Ekiel-Je\.zewska$^{\ast}$\textit{$^{a}$}
}

\begin{document}

\renewcommand{\thefootnote}{\fnsymbol{footnote}}
\renewcommand\footnoterule{\vspace*{1pt}
 \hrule width 3.5in height 0.4pt \vspace*{5pt}} 

\maketitle

\begin{abstract}

The settling of highly elastic non-Brownian closed fibres (called loops) under gravity in a viscous fluid is investigated numerically. 
The loops are represented using a bead-spring model with harmonic bending potential and finitely extensible nonlinear elastic (FENE) stretching potential. 
Numerical solutions to the Stokes equations are obtained with the use of HYDROMULTIPOLE numerical codes, which are based on the multipole method corrected for lubrication to calculate hydrodynamic interactions between spherical particles with high precision. 
Depending on the elasto-gravitation number $B$, a ratio of gravitation to bending forces, the loop approaches different attracting dynamical modes, as described by 
Gruziel-S\l omka \textit{et al.} (\textit{Soft Matt.},  vol.~15, 2019, pp.~7262-7274) 
with the use of the Rotne-Prager mobility of the elastic loop made of beads. 
Here, using a more precise method, we find and characterise a new mode, analyse typical timescales, velocities, and orientations of all the modes, compare them, and investigate their coexistence. 
We analyse the transitions (bifurcations) to a different mode at certain critical values of the elasto-gravitation number $B$. 
\\
\end{abstract}

\footnotetext{\textit{$^{a}$}~Institute of Fundamental Technological Research, Polish Academy of Sciences, ul. Pawi\'nskiego 5B, 02-106 Warsaw, Poland}

\footnotetext{$^{\ast}$~corresponding author: Maria L. Ekiel-Je\.zewska E-mail: mekiel@ippt.pan.pl}

\footnotetext{\dag~Author ORCIDs:
    Yevgen Melikhov, https://orcid.org/0000-0002-9787-5238; 
    Maria L. Ekiel-Je\.zewska, https://orcid.org/0000-0003-3134-460X.
}

\section{Introduction} \label{sec:intro}
Understanding sedimentation of deformable, complex-shaped objects is important for various biological systems of particles or cells, e.g., DNA, polymers, or red blood cells \cite{vologodskii_sedimentation_1998, lo_verso_end-functionalized_2008, peltomaki_sedimentation_2013, matsunaga_reorientation_2016, waszkiewicz_dna_2023}. 
The gravitational settling of particles of different shapes at the Reynolds number much smaller than unity has been of interest for a long time. The dynamics of various rigid objects have been studied, including conglomerates \cite{cichocki1995stokes}, trumbbells \cite{ekiel2009hydrodynamic}, slender ribbons \cite{koens2017analytical}, helical ribbons \cite{huseby2024helical}, helices \cite{palusa2018sedimentation}, propellers \cite{makino_sedimentation_2003}, disks \cite{chajwa_kepler_2019}, bent disks \cite{miara2024dynamics, vaquero2024u},  M\"obius bands \cite{moreno_sedimentation_2024}, and particles of general shapes \cite{witten_review_2020, goldfriend2015hydrodynamic, goldfriend2016hydrodynamic, joshi_fluttering_2024}. 
Depending on the shape, lateral motion, helical trajectories, periodic or quasi-periodic oscillations of singlets, and different patterns of hydrodynamic interactions within pairs have been observed.

For elastic objects, the motion can be even more complex, owing to time-dependent shapes. 
Typical patterns of evolution have been studied for varied elasticity and different shapes, such as filaments \cite{xu_deformation_1994, schlagberger_orientation_2005, cosentino_lagomarsino_hydrodynamic_2005, manghi2006hydrodynamic, llopis_sedimentation_2007, li_sedimentation_2013, saggiorato2015conformations, shojaei_sedimentation_2015, bukowicki_sedimenting_2019, du_roure_dynamics_2019, shashank_dynamics_2023, melikhov_attracting_2024}, dumbbells \cite{bukowicki2015dynamics}, trumbbells \cite{bukowicki_different_2018}, sheets \cite{miara2022sedimentation,yu2024free}, loops \cite{gruziel-slomka_stokesian_2019, waszkiewicz} and knots \cite{gruziel}.

Until now, most of papers on sedimenting deformable objects focused on the dynamics of moderately elastic filaments \cite{xu_deformation_1994, schlagberger_orientation_2005, cosentino_lagomarsino_hydrodynamic_2005, manghi2006hydrodynamic, llopis_sedimentation_2007, li_sedimentation_2013, shojaei_sedimentation_2015, bukowicki_sedimenting_2019, du_roure_dynamics_2019}. 
The main finding was the existence of a stable, stationary, planar, vertical configuration. The dependence of its U-like shape on the bending stiffness was determined. Recently, very different, rich dynamics of highly elastic filaments have been reported \cite{saggiorato2015conformations, shashank_dynamics_2023, melikhov_attracting_2024}. 
In these studies, the ends of the filament can move relative to each other. 
However, it is also interesting to investigate the dynamics of highly elastic loops. 

This work is focused on the analysis of different attracting dynamical modes of highly elastic loops settling under gravity in a viscous fluid at the Reynolds number much smaller than unity. We extend the previous results \cite{gruziel-slomka_stokesian_2019,waszkiewicz} by finding a new mode and analysing all the modes, discussing the characteristic time scales \& velocities, and investigating bifurcations between the modes at critical values of the bending stiffness. 

The plan of the paper is as follows. The theoretical model, its numerical implementation, and its parameters are presented in \textsection\thinspace \ref{sec:met}. 
Properties of the dynamical modes reached from an inclined planar and non-planar initial 
configurations for different values of the elasto-gravitation number are analysed in \textsection\thinspace \ref{ri} and \textsection\thinspace \ref{sec:bi}, respectively. Characteristic time scales and loop velocities for different attracting modes are analysed in \textsection\thinspace \ref{sec:time}.
Discussion and conclusions are presented in \textsection\thinspace \ref{sec:di}.

\section{Methodology}\label{sec:met}

\subsection{%
Model of elastic loops and their dynamics}

The bead-spring model is used to represent an elastic fibre.  %
The fibre %
is closed and it forms a loop. It consists of $N$ identical non-deformable spherical beads $N$ of %
diameter $d$. Position of the centre of the \textit{i}-th bead is denoted as %
$\bm{r}_i$, for $i=1,...,N$. 
Consecutive beads interact with each other by 
the finitely extensible nonlinear elastic
(FENE) spring potential energy \cite{warner_kinetic_1972, bird_dynamics_1977}, 
\begin{equation}
    U^{S} = - 
    \frac{1}{2} k (l_{0}-d)^{2} \sum_{i=1}^{N}\ln\left[ {1 - \left( \frac{l_{0}-l_{i}}{l_{0}-d} \right)^{2}} \right]. 
\label{US}\end{equation}
where 
$l_i=|\bm{r}_{i+1}-\bm{r}_{i}|$ for $i=1,...,N-1$, $l_N=|\bm{r}_{1}-\bm{r}_{N}|$,
$l_{0}$ is the equilibrium distance between beads centres, and
$k$ is a spring constant. %

The FENE spring potential allows for precise treatment of the dynamics of very close bead surfaces, preventing spurious overlaps. 
The choice of a small value %
$l_{0}=1.01d$ leads to %
small time-dependent gaps between the surfaces of the consecutive beads. %
Therefore, the loops are almost inextensible. %

In addition, there are also bending forces.
It is assumed that each triplet of the consecutive beads is straight at the elastic equilibrium, and it harmonically resists bending, leading to the following bending potential energy of the whole loop, 
\begin{equation}
    U^{B}= \sum_{j=1}^{N} \frac{A}{2 l_{0}} \beta_{j}^2,\label{UB}
\end{equation}
where
$A$ is the bending stiffness, and 
$\beta_{i}$ is the bending angle between the consecutive bonds, with %
$\cos \beta_i\!=\!({\bm{r}}_i\!-\!{\bm{r}}_{i-1})\cdot({\bm{r}}_{i+1}\!-\!{\bm{r}}_{i})/(l_{i-1} l_{i})$
for $i=2,...,N-1$, $\cos \beta_1=(\bm{r}_{1}-\bm{r}_{N})\cdot(\bm{r}_{2}-\bm{r}_{1})/(l_{N} l_{1})$, and $\cos \beta_N=(\bm{r}_{N}-\bm{r}_{N-1})\cdot(\bm{r}_{1}-\bm{r}_{N})/(l_{N-1} l_{N})$. 
For highly elastic fibres, the harmonic bending potential energy
\cite{mackerell1998all,storm2003theory,frenkel2023understanding}, used in this work, is more realistic than %
the Kratky-Porod potential energy %
\cite{schlagberger_orientation_2005,manghi2006hydrodynamic,marchetti_deformation_2018,llopis_sedimentation_2007,saggiorato2015conformations}. The reason is that %
for larger bending angles (as it happens for highly elastic fibres), the Kratky-Porod potential may lead to spurious dynamics \cite{bukowicki_different_2018}. %

The bending stiffness $A$ is proportional to the Young's modulus $E_{Y}$ by the model of an elastic cylinder of diameter $d$ \cite{bukowicki_different_2018, bukowicki_sedimenting_2019},
\begin{equation}
    A = \frac{E_Y \pi d^4}{64}. 
\end{equation} 
The same elastic  model is applied to 
 the spring constant $k$, and therefore 
 the spring constant $k$ is linked to the bending stiffness $A$ by the relation \cite{bukowicki_different_2018}.
\begin{equation}
    A = 
    \frac{k d^{2} l_{0}}{ 16}.
\end{equation} 

The potential energies \eqref{US}-\eqref{UB} result in the following expression for the elastic force %
onto the $j$-th bead, for $j=1,...,N$:
\begin{equation}
    \bm{F}_j^e = - \frac{\partial \left( U^{S} +  U^{B} \right)}{\partial \bm{r}_j}.
\end{equation}

There is also a constant gravitational force, corrected for buoyancy,  that acts on each bead $j$ along the $z$-axis: 
\begin{equation}
    \bm{F}^g = - \frac{G}{N} \cdot \bm{ \hat{e} }_z.
\end{equation}
Here, $G$ is the total gravitational force on the whole loop,
and 
$\bm{ \hat{e} }_z$ is the unit vector along the $z$-axis.

We restrict to the systems with the Reynolds number much smaller than unity. Therefore, the fluid flow obeys the Stokes equations, and the bead velocities, $\dot{\bm{r}}_{i}$, depend linearly on the elastic and gravitational forces exerted on the beads $j$. The dynamics of the positions of the bead centres  $\bm{r}_i$, $i=1,...,N$, satisfy the set of the first-order ordinary differential equations, 
\begin{equation}
    \dot{\bm{r}}_{i} = \sum_{j=1}^{N} \bm{\mu}_{ij}(\bm{r}_1,...,\bm{r}_N) \cdot \left( \bm{F}_j^e + \bm{F}^g \right).\label{Sd}
\end{equation}
The 3x3 mobility matrices $\bm{\mu}_{ij}(\bm{r}_1,...,\bm{r}_N)$, for $i,j=1,...,N$, depend on the time-dependent positions of all the bead centres. They are calculated %
by the multipole expansion of the solutions to the Stokes equations, corrected for lubrication to speed up the expansion convergence \cite{Felderhof1988,Cichocki1994,cichocki_lubrication_1999}.

In the following, we choose $G/N$ as the force unit, and adopt the following units for the length, time, and translational and rotational velocities:
\begin{equation}
  d, \quad 
  \tau_{b}= \frac{\pi \eta d^2N}{G}, \quad 
  v_{b}=\frac{d}{\tau_{b}}=\frac{G}{\pi \eta d N}, \quad
  \omega_{b}=\frac{1}{\tau_{b}}=\frac{G}{\pi \eta d^2N}.\label{normal}
\end{equation}
From now on, we redefine the symbols used previously to mean the corresponding dimensionless quantities. 

\subsection{Simulations, parameters and variables}

The precise numerical code {\sc Hydromultipole} \cite{cichocki_lubrication_1999,ekiel-jezewska_precise_2009}, based on multipole expansion of solutions to Stokes equations, corrected for lubrication \cite{cichocki_lubrication_1999}, is used to evaluate the mobility matrices $\bm{\mu}_{ij}(\bm{r}_1,...,\bm{r}_N)$ and solve the dynamics in \eqref{Sd}. 
The multipole truncation order $L=2$ is taken in these computations. 
The majority of the simulations were performed until $t=88000$. 
A few selected simulations were run for longer times.

In the simulations, the number of beads in the loop is fixed, $N=36$. 
Following Refs.~\cite{llopis_sedimentation_2007, cosentino_lagomarsino_hydrodynamic_2005, saggiorato2015conformations, marchetti_deformation_2018, bukowicki_different_2018, bukowicki_sedimenting_2019}, we introduce in this work the elasto-gravitation number, $B$, that is a ratio of the gravitational and bending forces acting on the loop,
\begin{equation}
   B  = \dfrac{N^2 d^2 G}{A}. 
   \label{Bdef}
\end{equation}
The value of $B$ was varied in the range $1000 \le B \le 40000$. 
Within the considered range of values of the elasto-gravitation number, the moderately elastic loops, with smaller values of $B$, deform from a circle only slightly. 
However, highly elastic loops, with larger values of $B$, deform significantly out of a plane.

To study the time evolution of the loop orientation, we evaluate the time-dependent gyration tensor \cite{mattice1994conformational},
\begin{equation}
    {S}_{\alpha \beta} =
    \frac{1}{N}\sum\limits_{j=1}^{N} r'_{j\alpha} r'_{j\beta},
    \label{gt}
\end{equation}
where $\alpha = x, y, z,\;\beta = x, y, z$ and  ${\bm{r}}'_j=(r'_{jx},r'_{jy},r'_{jz})$ is the position of $j$-th bead centre in the centre-of-mass reference frame, ${\bm{r}}'_j={\bm{r}}_j-\bm{r}_{CM}$, with the centre-of-mass position 
\begin{equation}
    {\bm{r}}_{CM}\equiv \frac{1}{N}\sum_{i=1}^N \bm{r}_i=(x_{CM},y_{CM},z_{CM}).
\end{equation}
We find the eigenvalues and eigenvectors of the gyration tensor defined in \eqref{gt}. 
We select the unit eigenvector associated with the smallest eigenvalue and denote it as $\bm{n}$. 
The unit vector $\bm{n}$ is used to provide information about the orientation of the loop.
For example, in the case of a circular flat loop with its centre at the origin of the coordinate system, the unit vector $\bm{n}$ is perpendicular to the plane of the loop, and it lies on the loop rotational symmetry axis.

Expressing $\bm{n}$ in spherical coordinates with the polar angle $\theta$ that it makes with the vertical axis $z$ (anti-parallel to gravity), and the azimuthal angle $\phi$ that its projection onto the $xy$-plane makes with the $x$-axis, 
\begin{equation}
    \bm{n} = \begin{pmatrix}
              \sin \theta \cos \phi, \sin \theta  \sin \phi,  \cos \theta
              \end{pmatrix}
  \end{equation}
we will monitor dependencies of $\theta$ and $\phi$ on time, and we will use them to help us to characterize the dynamical modes.

\section{The dynamical modes reached from an inclined planar initial configuration}
\label{ri}

\subsection{Outline}

In this section, we present the majority of our simulations. 
We focus on a flat inclined circle as the initial configuration, similarly as in \cite{gruziel-slomka_stokesian_2019}. 
Typically, $\theta(t=0)=16^{\circ}$ for highly elastic loops, and $\theta(t=0)=80^{\circ}$ for some moderately elastic loops. 
A few simulations performed with curved initial configurations will be presented 
in~\textsection\thinspace \ref{sec:bi}.

We find numerically that a highly elastic loop, after a relatively long time, reaches a certain dynamical mode and remains in this mode for a very long time until the end of the simulations. 
For a given initial configuration, the type of the mode reached depends on a value of the elasto-gravitation number $B$. 
The observed dynamical modes are presented in the Supplementary Movies. 
The evolution of the shapes in the vertical, tilted, frozen rotating, tank-treading (TT), swinging and flapping modes observed in this work is qualitatively similar to that reported earlier in \cite{gruziel-slomka_stokesian_2019}, where a numerical study was performed using the Rotne-Prager method. 
Two additional modes are described in this work: rocking and gyrating-rocking-tank-treading (GRTT). 
The rocking mode might be the same as the tilted swinging mode in \cite{gruziel-slomka_stokesian_2019}, but the shape evolution of the tilted swinging mode was not provided there. (The evolution of shapes in the rocking mode is very different from that in the swinging mode.)
\begin{figure}
    \centering 
    \includegraphics[width=0.91\textwidth]{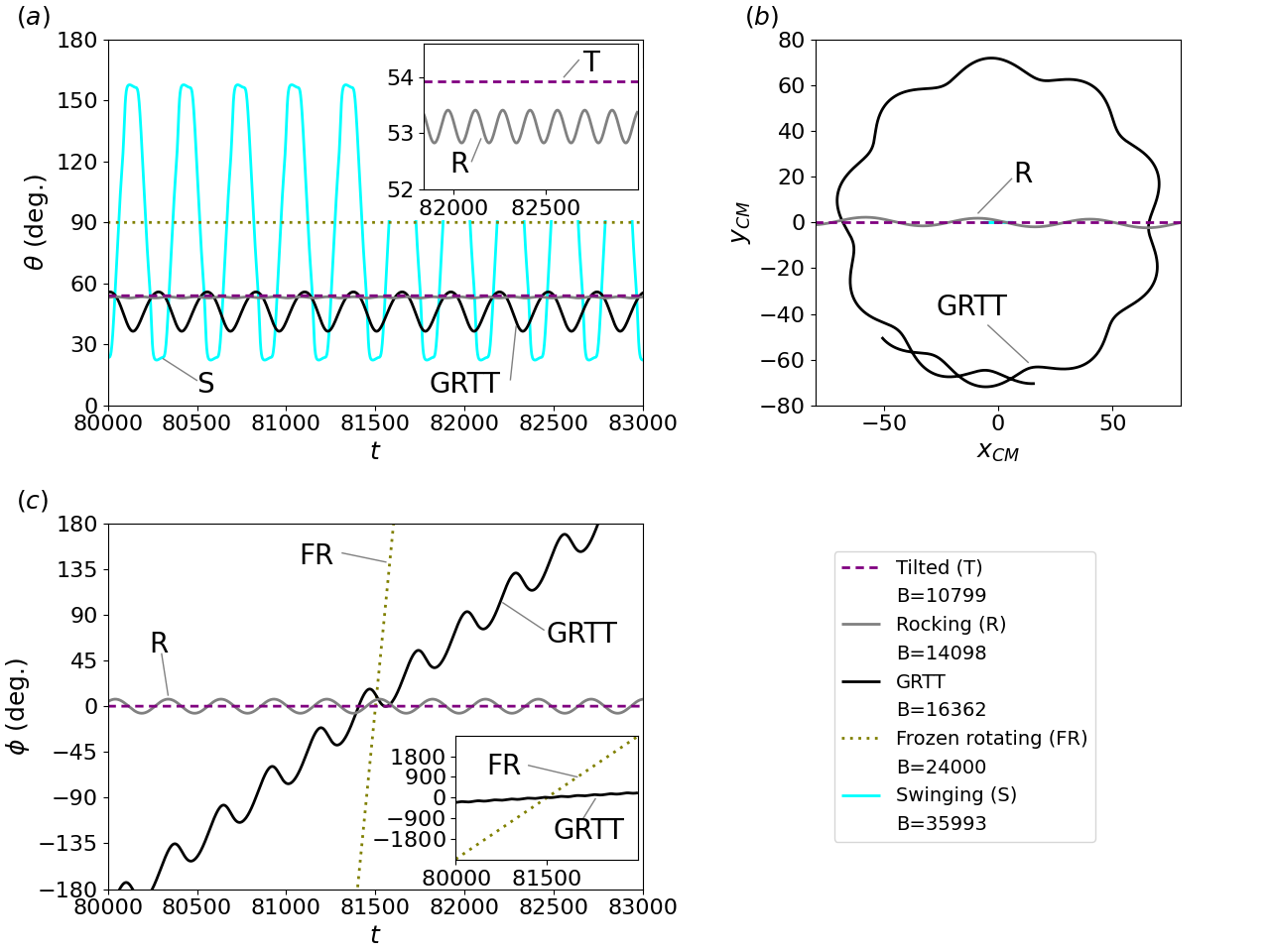}
    \vspace{-0.2cm}
    \caption{
    Different attracting dynamical modes: the time dependence of 
    (\textit{a})~polar angle $\theta$ and 
    (\textit{c})~azimuthal angle $\phi$, together with 
    (\textit{b})~horizontal projections of the centre-of-mass trajectories.  
    In (\textit{b}), the ranges of times are: $963$ for the tilted, $968$ for the rocking, and $3000$ for the GRTT modes. 
    In (\textit{b}), for the swinging mode, the centre-of-mass of the loop oscillates on the line $y_{CM} \!=\! 0$ between $-7 \le x_{CM} \le 7$, which it covers during time period of $303$; for the frozen rotating mode, $x_{CM}\! = \!0$ and $y_{CM}\! = \!0$. 
    }
\label{fig:tilted_angles_vs_time_diiff_modes}
\end{figure}

For different modes, we analyse the time dependence of the loop orientation, $\theta$, $\phi$, and its centre-of-mass motion, $(x_{CM},y_{CM},z_{CM})$.
The exemplary comparison of $\theta(t)$, $\phi(t)$, and $y_{CM}(x_{CM})$ for different attracting modes is presented in figure~\ref{fig:tilted_angles_vs_time_diiff_modes}. 

For the vertical mode, $\theta=90\degree$, $\phi=0\degree$ and $x_{CM}\! = \!y_{CM}\! = \!0$. 
Figure~\ref{fig:tilted_angles_vs_time_diiff_modes} does not include the tank-treading, flapping, and irregular modes for which significant out-of-plane loop deformations are observed.

In the next sections, we determine the typical properties of each mode, including the analysis of the characteristic time scales. 
We also identify critical values of $B$ for the transitions between different modes.

\subsection{Vertical and tilted modes}

\begin{figure}
    \centering
    \includegraphics[width=0.46\textwidth]{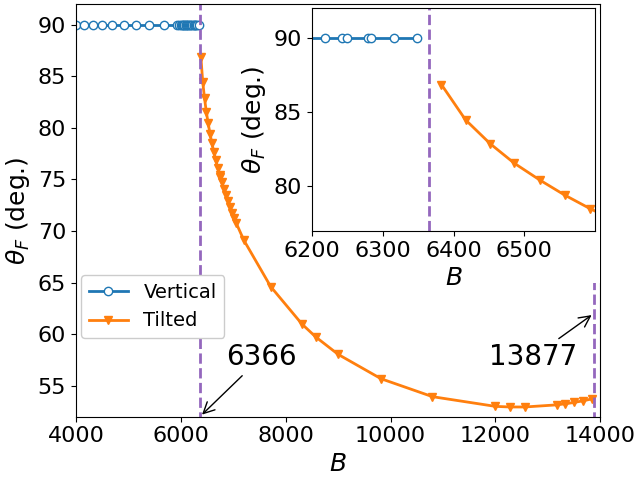}
    \caption{
    The final polar angle $\theta_{F}$ as a function of the elasto-gravitation number $B$ for the tilted and vertical modes.
    Symbols correspond to the numerical data and are connected by solid lines to guide the eye. 
    }
    \label{fig:theta_tilted_vertical}
\end{figure}

A moderately elastic loop (i.e.,  with lower values $1000 \leq B \leq 13847$ of the elasto-gravitation number), initially planar, circular, and inclined, later deforms a bit \cite{gruziel-slomka_stokesian_2019}, and after a long time attains a fixed shape with two or one vertical planes of symmetry, in a vertical or tilted mode, respectively. 
In the vertical mode,  the loop shape is restricted to a vertical plane, but it is not circular \cite{gruziel-slomka_stokesian_2019}.
The corresponding loop shapes are shown in Movies 1 and 2.
For a given value of $B$, such a mode is characterized by a single, constant in time,  value of the polar angle \begin{equation} \theta(t)=\theta_{F},\end{equation} and a constant in time value of the azimuthal angle; in the chosen coordinate system, \begin{equation}\phi(t)=0. \end{equation}
In the vertical mode, $\theta_{F}=90 \degree$, and in the tilted mode, $\theta_{F} < 90\degree$.
The dependence of the final angle $\theta_{F}$ on the elasto-gravitation number $B$ is shown in figure~\ref{fig:theta_tilted_vertical} (analogous to figure~16 in \cite{gruziel-slomka_stokesian_2019}). 
The loop inclination in the tilted mode results in a drift of its centre of mass in the $x$-direction, as illustrated in figure~\ref{fig:tilted_angles_vs_time_diiff_modes}(c).

\begin{figure}
    \centering
    \includegraphics[width=0.91\textwidth]{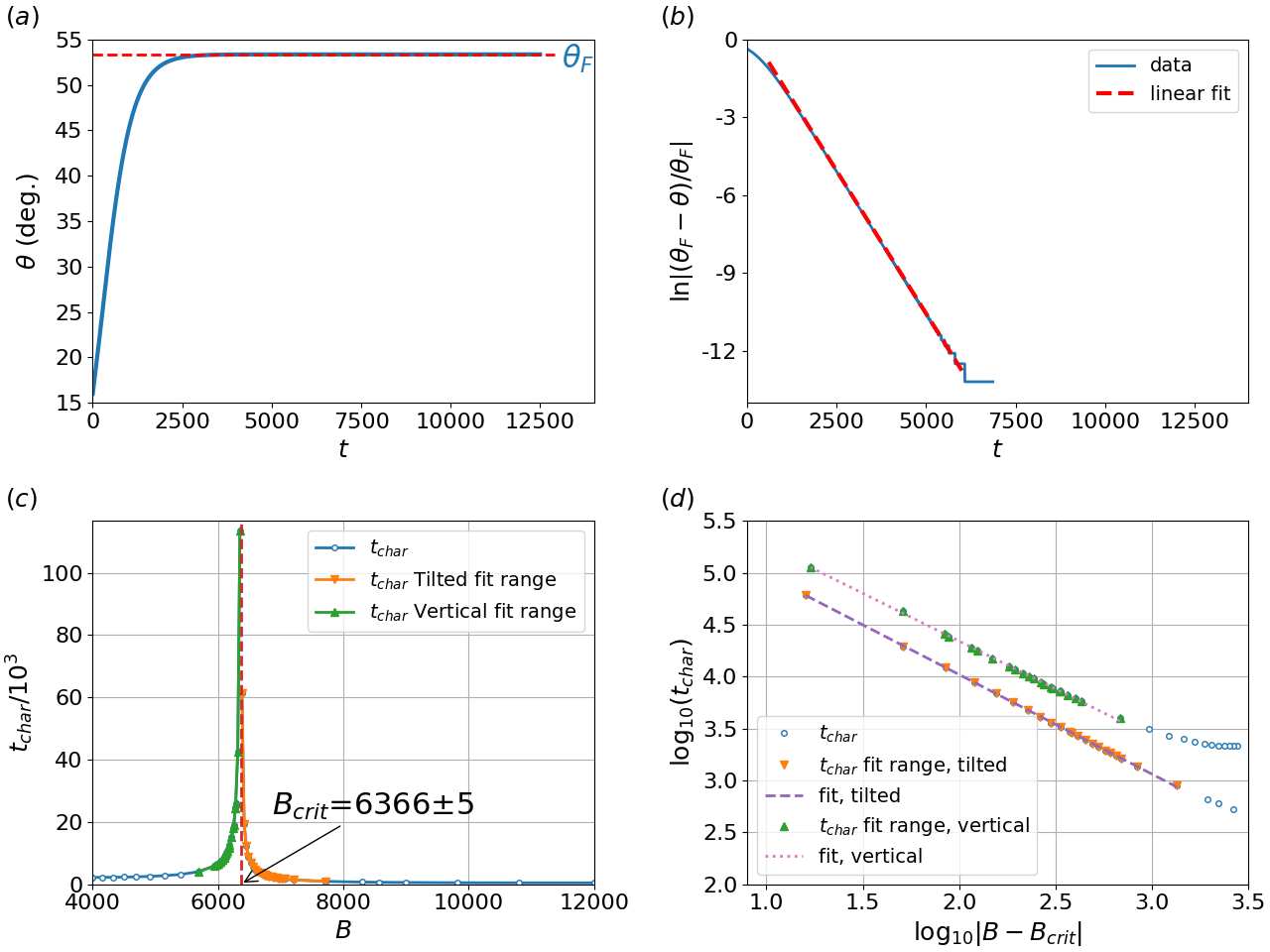}
    \vspace{-0.2cm}
    \caption{
    The exponential approach of the polar angle $\theta(t)$ to the final angle $\theta_{F}$ for vertical and tilted modes with different values of $B$ can be used to estimate $B_{crit}$ between both modes. 
    (\textit{a})~$\theta(t)$ for $B=13501$ with $\theta_{F}=53.37\degree$.
    (\textit{b})~$\mathrm{ln} \left| (\theta_{F} - \theta )/ \theta_{F}  \right|$ is well approximated by the fit \eqref{eq:theta_tilted_vertical} with $t_{char}=453$. 
    (\textit{c})~The characteristic time $t_{char}$ of the fit \eqref{eq:theta_tilted_vertical} as a function of the elasto-gravitation number $B$. 
    (\textit{d})~Log-log plot confirms a power law dependence of $t_{char}$ on $|B-B_{crit}|$ near the critical value $B_{crit}$. 
    }
    \label{fig:tchar_tilted_vertical}
\end{figure}

The transition value $B_{crit}$ between the vertical and tilted modes, marked in figure~\ref{fig:theta_tilted_vertical}, has been determined by analyzing how these modes are reached from the initially inclined circular configuration. 
For both modes, the inclination angle $\theta$ monotonically increases with time, and the exponential law can fit this relation, as shown in figure~\ref{fig:tchar_tilted_vertical}(a),
\begin{equation}
  \theta = \theta_{F} \cdot ( 1 - e^{-t/t_{char}} ),
  \label{eq:theta_tilted_vertical}
\end{equation}
with characteristic time $t_{char}>0$ depending on the elasto-gravitation number $B$ and different for the tilted and vertical modes.
Numerical simulations show that when $B$ is approaching $B_{crit}$ from either side, the characteristic time $t_{char}$ approaches infinity, as presented in figure~\ref{fig:tchar_tilted_vertical}(b). Therefore, the power-law increase is expected,
\begin{equation}
  t_{char} = t_{0} \cdot |B-B_{crit}|^{-p}.
  \label{eq:tchar_exp}
\end{equation}
In figure~\ref{fig:tchar_tilted_vertical}(c), the fit \eqref{eq:tchar_exp} is plotted together with the numerical data. 
The fitted parameters are $B_{crit} = 6366$, $(t_0,p)=(1.528 \cdot 10^6, 0.922)$ for the vertical mode and $(t_0,p)=(0.863 \cdot 10^6, 0.959)$  for the tilted mode.

The evolution of the loop's centre of mass in the vertical and tilted modes can be expressed in the following simple form, 
\begin{equation}
  \begin{cases}
    x_{CM}(t)=x_{c} + %
    v_{CM,x} t, \\
    y_{CM}(t)=y_{c}, \\
    z_{CM}(t)=-v_{CM,z} t,
  \end{cases}
  \label{eq:vert_tilt}
\end{equation}
where $x_{c}$, $y_{c}$, $v_{CM,x} \ge 0$ and $v_{CM,z}\ge 0$ are constants, and in the case of vertical mode, $v_{CM,x}=0$.
The sedimentation velocity $v_{CM,z}$ in the vertical mode is almost independent of $B$: $1.631 \lesssim v_{CM,z} \lesssim 1.635$.
On the other hand, the velocity $v_{CM,x}$ increases and $v_{CM,z}$ decreases with the increase of $B\le 12000$, what is caused by the decrease of the inclination angle $\theta$, shown in figure~\ref{fig:theta_tilted_vertical}, as will be discussed in \textsection\thinspace \ref{sec:di}.

\subsection{Rocking and gyrating-rocking-tank-treading modes}

In our numerical simulations, elastic loops with the elasto-gravitation number $13882 \leq B \leq 14211$ end up in the rocking mode, illustrated in Movie 3. 
For slightly larger values, $14248 \leq B \leq 17422$, a gyrating-rocking-tank-treading mode is reached, as shown in Movie 4. 
The common feature of these modes is the existence of periodic in-time oscillations of both orientation angles $\theta$ and $\phi$, as shown in figure~\ref{fig:tilted_angles_vs_time_diiff_modes}(a)-(b). 
We first analyse the properties of the rocking mode, and then we describe the gyrating-rocking-tank-treading mode.

\begin{figure}
    \centering
    \includegraphics[width=0.91\textwidth]{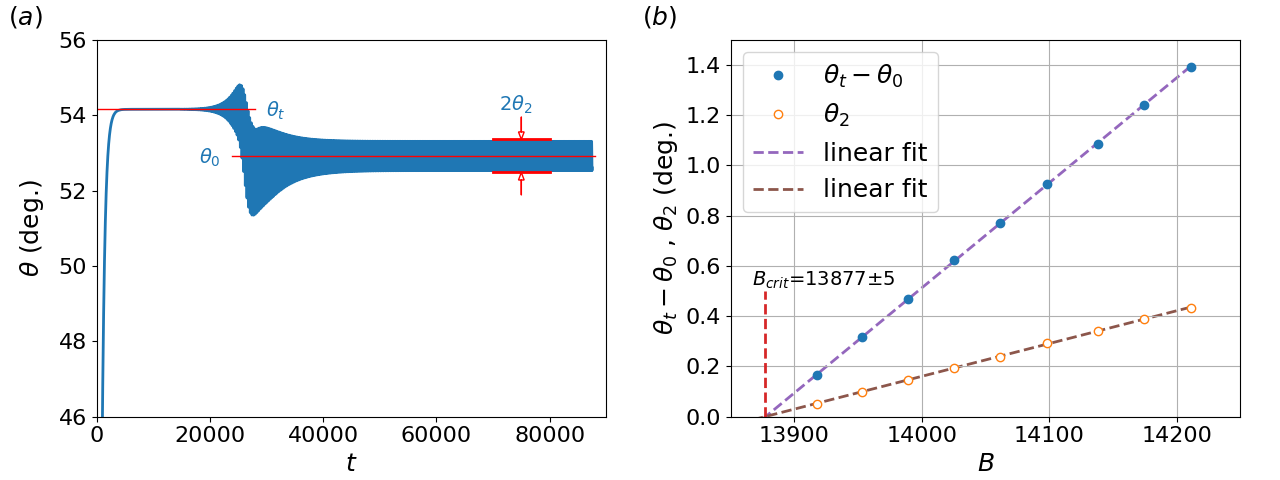}
    \vspace{-0.2cm}
    \caption{
    The evolution of $\theta$ for different values of $B$ is used to determine $B_{crit}$ for the transition between tilted and rocking modes. 
    (\textit{a})~$\theta(t)$ for  $B=14174$. %
    For $4000 \lesssim t \lesssim 18000$, the loop seems to be in the titled mode with $\theta_{t} \approx 54.16\degree$. 
    But later, it destabilizes. 
    The periodic rocking motion is observed after the transition phase is finished at $t \approx 55000$.
    (\textit{b})~$\theta_t-\theta_0$ and $\theta_2$ are well-fitted by linear functions of $B$; they vanish at approximately the same value of $B_{crit} = 13877$, estimated as the transition between tilted and rocking modes.     
    }
    \label{fig:Rocking_Bcrit}
\end{figure}

When the elasto-gravitation number $B$ increases above the values typical for the tilted mode, the loop ends in a rocking mode instead of the tilted one. 
The appearance of this mode is illustrated in figure~\ref{fig:Rocking_Bcrit}(a). 
In the first stage of the evolution, the loop seems to reach the tilted mode with a certain inclination angle $\theta_{t}$. 
However, after a relatively long time, the inclination angle $\theta(t)$ decreases, and starts oscillating around a lower value $\theta_0$. 
The periodic oscillations of the angle $\theta$, with a constant amplitude $\theta_2$ and the average $\theta_0\ne 90\degree$, are typical for the rocking mode. 

The amplitude $\theta_{2}$ of the rocking oscillations and the difference $\theta_t-\theta_0$ increase linearly with an increase of the elasto-gravitation number $B$ as seen in figure~\ref{fig:Rocking_Bcrit}(b). 
Linear fits to the simulation data predict that $\theta_{2}$ and $\theta_t-\theta_0$ vanish at approximately the same value $B=B_{crit} \approx 13877$. 
This observation indicates that the rocking mode ceases to exist when $B$ decreases to $B_{crit}$.

\begin{figure}
    \centering
    \includegraphics[width=0.46\textwidth]{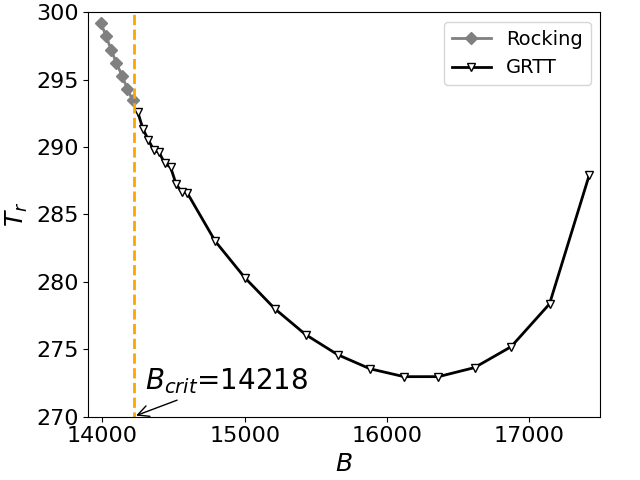}
    \vspace{-0.2cm}
    \caption{
    Dependence of the period $T_r$ of the oscillations on elasto-gravitation number $B$ in the rocking mode and in the gyrating-rocking-tank-treading mode. 
    }
    \label{fig:period_rocking}
\end{figure}

In the rocking mode, the period $T_r={2\pi}/{\omega_r}$ of the azimuthal angle oscillations is twice as large as the period of the polar angle oscillations, as visible in figure~\ref{fig:tilted_angles_vs_time_diiff_modes}. 
For the rocking mode, the period $T_{r}$ %
decreases gradually with an increase of $B$, as shown in figure~\ref{fig:period_rocking} (blue triangles). 
The oscillations of $\theta$ are around the averaged inclination angle $\theta_0\ne 90\degree$. 
We choose such a reference frame that the oscillations of $\phi$ are around the average value equal to zero. 
The time dependence of $\theta$ and $\phi$ can be approximated as
\begin{equation}
  \begin{cases}
    \theta(t) = \theta_{0} + 
                \theta_{2}  \sin \left( 2 \omega_{r}  t + \psi_{\theta} \right),
    \\
    \phi(t)   = \mkern 35mu 
                \phi_{1} \sin \left( \mkern 9mu \omega_{r}  t + \psi_{\phi} \right),
  \end{cases}
  \label{eq:rocking_angles}
\end{equation}
where $\psi_{\theta}$ and $\psi_{\phi}$ are the phase shifts, and the amplitudes $\theta_{2}$ and $\phi_{1}$ dependent on $B$. 
In figure~\ref{fig:fit_rocking_angles} in Appendix~\ref{app:fits}, the approximation from \eqref{eq:rocking_angles} is shown to fit well the numerical data for an exemplary value of $B$. 
The amplitudes $\theta_{2}$ and $\phi_{1}$ are small. %
The largest value of $\phi_1$, i.e., $\phi_{1}=8.52\degree$ is attained for $B=14211$. 

In contrast, $\theta_0$ is relatively large, as shown in figure~\ref{fig:theta_tilted_vertical}.
The averaged inclination of the loop is associated with an averaged horizontal drift of its centre of mass along the $x$-axis, as illustrated in figure~\ref{fig:tilted_angles_vs_time_diiff_modes}(c).

The motion of the loop's centre of mass in the rocking mode is a superposition of translation with a constant velocity and periodic oscillations. 
It can be approximated as 
\begin{equation}
  \begin{cases}
    x_{CM}=v_{x,c}\,  t + 
        A_{r,x}  \cos \left(    2 \omega_{r}  t + \psi_{r,x} \right),
        \\
    y_{CM}= %
        \mkern 57mu 
        A_{r,y}  \cos \left( \mkern 9mu \omega_{r}  t + \psi_{r,y} \right),
        \\
    z_{CM}=v_{z,c}\,  t + 
        A_{r,z}  \cos \left(    2   \omega_{r}  t + \psi_{r,z} \right).
  \end{cases}
  \label{eq:rocking_cm}
\end{equation}
where $\psi_{r,x}$, $\psi_{r,y}$ and $\psi_{r,z}$ are the phase shifts, 
velocities $v_{x,c}$ and $v_{z,c}$ 
depend on $B$, and amplitudes $A_{r,x}$, $A_{r,y}$ and $A_{r,z}$ decrease with a decrease of $B$. 
In figure~\ref{fig:fit_rocking_cm} in Appendix~\ref{app:fits}, the approximation from \eqref{eq:rocking_cm} is shown to fit well the numerical data for an exemplary value of~$B$.

Gyrating-rocking-tank-treading (GRTT) mode is observed for moderately elastic loops with elasto-gravitation number $B$  in the range $14248 \leq B \leq 17422$. 
As a part of this name indicates, the loop `gyrates' -- it rotates with a constant angular velocity $\omega_g$ around a vertical axis, and also performs `rocking' oscillations with the frequency $\omega_r$, similar to in the rocking mode. 
The polar angle is a periodic function of time, with the period $T_r=2\pi/\omega_r$. 
The azimuthal angle is a superposition of the linear growth in time at the rate equal to $\omega_g$ and a periodic function with the period $T_r$.  
The time-dependent angles $\theta(t)$ and $\phi(t)$ can be approximated as
\begin{equation}
  \begin{cases}
    \theta(t) = \theta_{0} \; \; + 
                \theta_{1} \sin \left( \omega_{r} t + \psi_{\theta1} \right) + 
                \theta_{2} \sin \left( 2  \omega_{r}  t + \psi_{\theta2} \right),
    \\
    \phi(t)   = %
    \omega_g  t + 
                \phi_{1}  \sin \left( \omega_{r}  t + \psi_{\phi1} \right) + 
                \phi_{2}  \sin \left( 2  \omega_{r}  t + \psi_{\phi2} \right),
  \end{cases}
  \label{eq:GRTT_angles}
\end{equation}
where the parameters depend on elasto-gravitation number $B$.
In general, two frequencies are needed in \eqref{eq:GRTT_angles}.

The evolution of the loop's centre of mass as a function of time is more complex. 
The horizontal projection of the centre-of-mass trajectory, shown in figure~\ref{fig:tilted_angles_vs_time_diiff_modes}(c), involves two characteristic frequencies, $\omega_g$ and $\omega_r$, with in general irrational ratio. 
Therefore, the centre-of-mass motion, in general, is quasi-periodic. 
It can be approximated by the following equations,
\begin{equation}\hspace{-0.3cm}
  \begin{cases}
    x_{CM}=x_{c} + 
        A_{g}  \cos \left( \omega_{g}  t \right) +
        A_{r,xy}  \left[
        \cos \left( ( \omega_{r} - \omega_{g} )  t + \phi_{r,xy_{1}} \right) + 
        \cos \left( ( \omega_{r} + \omega_{g} )  t + \phi_{r,xy_{2}} \right)
        \right],
        \\
    y_{CM}=y_{c} + 
        A_{g}  \sin \left( \omega_{g} t \right) +
        A_{r,xy}  \left[
        \sin \left( ( \omega_{r} - \omega_{g} )  t + \phi_{r,xy_{1}} \right) + 
        \sin \left( ( \omega_{r} + \omega_{g} )  t + \phi_{r,xy_{2}} \right)
        \right],
        \\
    z_{CM}=v_{z,c}  t + 
        A_{r,z}  \sin \left( \omega_{r}  t + \phi_{r,z} \right),
  \end{cases}
  \label{eq:grtt_cm}
\end{equation}
with the parameters dependent on $B$. %
$A_{g}$ is the amplitude (the averaged radius) of the gyrating motion and %
the amplitudes $A_{r,xy}$ %
and $A_{r,z}$ are associated with the oscillating `rocking' motion. 
The constants $x_{c}$ and $y_{c}$ are somewhat arbitrary. %
Their values are set at the end of a transition phase. 
The sedimentation velocity $v_{z,c}$ is constant in time. %
In figures~\ref{fig:fit_grtt_angles}-\ref{fig:fit_grtt_cm} in Appendix~\ref{app:fits}, the approximations from \eqref{eq:GRTT_angles}-\eqref{eq:grtt_cm} are shown to fit the numerical data well for an exemplary value of $B$. 

\begin{figure}
    \centering
    \includegraphics[width=0.91\textwidth]{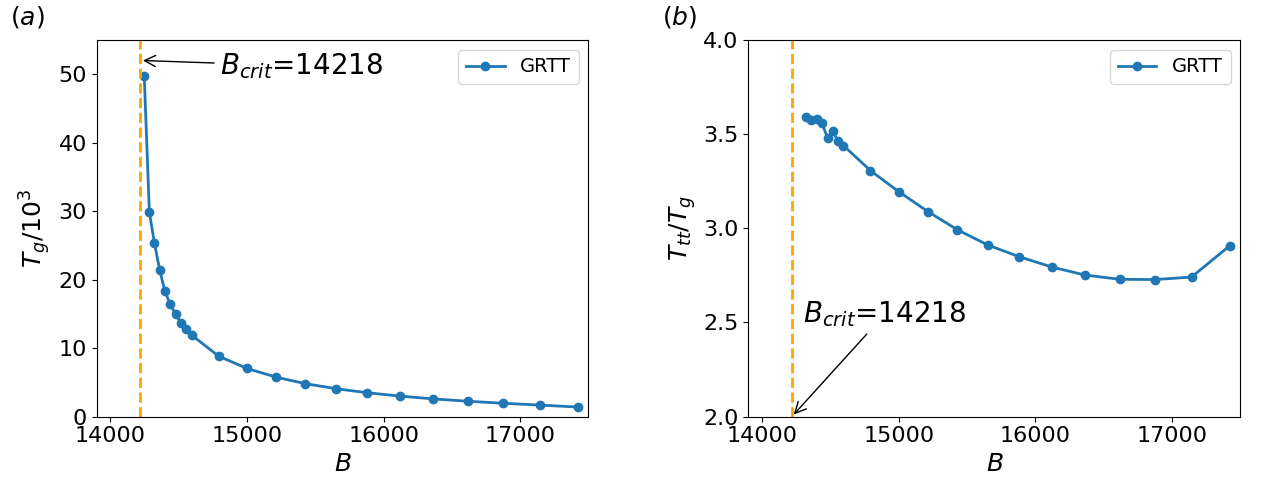}
    \vspace{-0.2cm}
    \caption{
    Dependence of the characteristic time scales of the GRTT motion on the elasto-gravitation number $B$:
    (\textit{a})~$T_g$,  
    (\textit{b})~$T_{tt}/T_g$. 
    }
    \label{fig:period_grtt_cm_gyration}
\end{figure}

The function $T_r(B)$ is shown in figure~\ref{fig:period_rocking}. 
The period of the oscillations in the GRTT mode smoothly extends $T_r(B)$ from the rocking mode for larger values of $B$. 
The characteristic time scale of gyration, $T_g=2\pi/\omega_g$, is plotted as a function of $B$ in figure~\ref{fig:period_grtt_cm_gyration}(a). 
It is clear that $T_g$ increases significantly with the decrease of $B$. 
We expect that $T_g$ might diverge at a critical value of $B_{crit}$ as 
\begin{equation}
   T_g \sim |B-B_{crit}|^{-\chi}\label{Tg} 
\end{equation}
that allowed us to identify $B_{crit} = 14218$ as the critical value of $B$ corresponding to the transition between the rocking and GRTT modes (see Appendix~\ref{app:GRTT_Bcrit} for details). 

\begin{figure}
    \centering
    \includegraphics[width=0.91\textwidth]{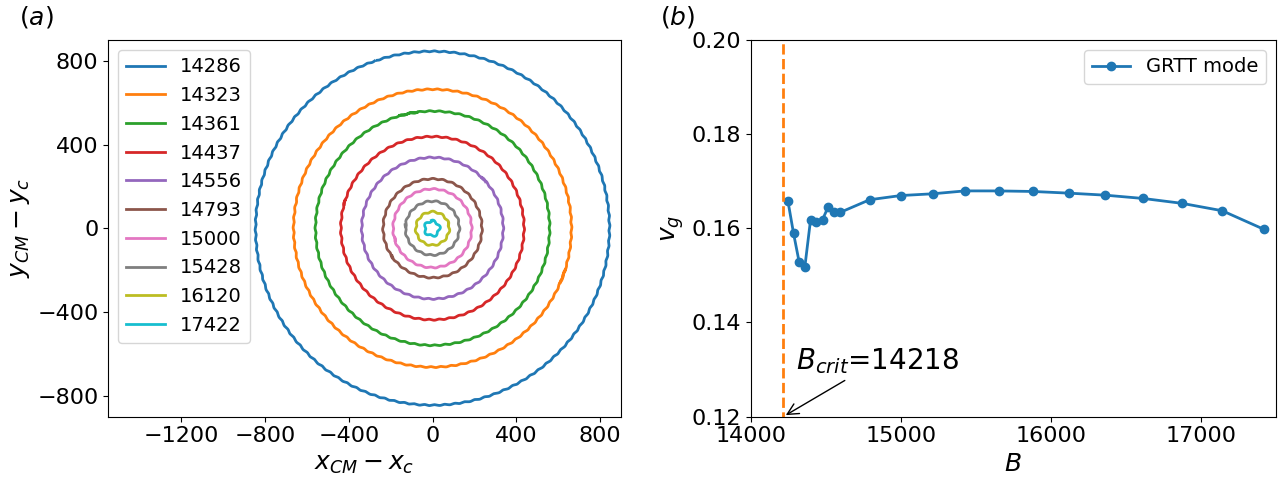} 
    \vspace{-0.2cm}
    \caption{
    In the GRTT mode, the averaged radius $A_r$ of the horizontal projection of the centre-of-mass trajectories increases with the decrease of $B$, as shown in (\textit{a}), while the gyration velocity is almost constant, as visible in (\textit{b}). 
   }
    \label{fig:radius_grtt_cm_gyration}
\end{figure}

The increase of $T_g$ is related to the increase of the amplitude (the averaged radius) $A_{g}$ of the gyrating motion. 
$T_g$ is proportional to $A_{g}$, with the approximately constant gyration velocity, $ v_g=A_g\,\omega_g \approx 0.16$, as shown in figure~\ref{fig:radius_grtt_cm_gyration}.

\begin{figure}
    \centering
    \includegraphics[width=0.91\textwidth]{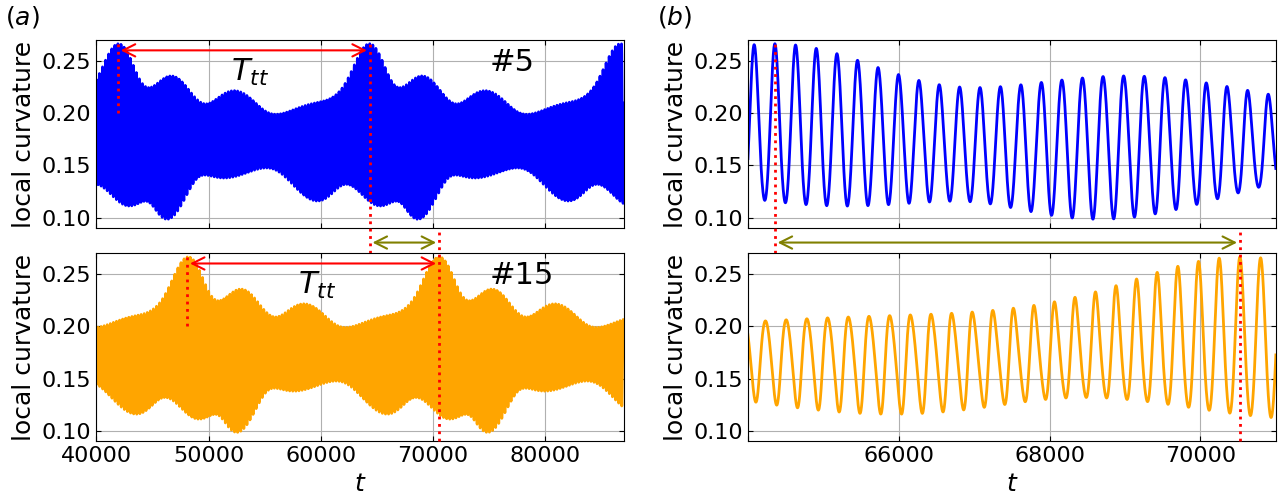}
    \vspace{-0.2cm}
    \caption{
    Time dependence of the local curvature for the 5\textsuperscript{th} and 15\textsuperscript{th} beads in the GRTT mode with $B=15000$ is described by approximately the same envelope function of a period $T_{tt}$, but shifted in time by $10T_{tt}/36$, as shown in (\textit{a}).
    Oscillations of the local curvature at a short-time scale $T_r$ due to the rocking motion are seen in (\textit{b}).
   }
    \label{fig:grtt_local_curv}
\end{figure}

So far, we have analysed only global features of the loop, such as the polar and azimuthal angles and the centre-of-mass coordinates or velocities. 
Some information on the loop shape can be provided by the time-dependent local curvature at each specific bead $i$, which is calculated as the inverse of the radius of a circle circumscribed on the centres of three consecutive beads $i-1,i,i+1$ \cite{slowicka_buckling_2022}. 
An example of the time-dependent curvature at two beads for $B=15000$ is shown in  figure~\ref{fig:grtt_local_curv}.  
The local curvature has approximately the same envelope function of time for both beads but with a time shift. 
The shift corresponds to the tank-treading-like motion of each bead along the loop shape.
This motion %
may be described as the beads undergoing a periodic alteration in their spatial position with respect to the centre of mass.
It bears a resemblance to the tank-treading motion in which the beads move along the fixed shape (see \textsection\thinspace \ref{sec:othermodes}).
Note that the overall shape of the loop in the GRTT mode is not fixed; the rocking oscillations take place.

The period $T_{tt}$ %
associated with the tank-treading-like motion can be also identified through an analysis of the time dependence of $z-z_{CM}$ for any given bead. 
Since beads undergo a periodic alteration in their spatial position with respect to the centre of mass, the quantity $z-z_{CM}$ will exhibit maxima and minima with a period of $T_{tt}$, similar to the local curvature. 

The time shift between the beads %
can be utilized to identify the value of $T_{tt}$ also when the time range of the observed %
GRTT mode is smaller than the period $T_{tt}$, which is particularly relevant in cases with smaller values of $B$, specifically when $B \le 14556$.

\subsection{Frozen rotating mode}\label{sec:frozenrotatmode}

\begin{figure}
    \centering
    \includegraphics[width=0.46\textwidth]{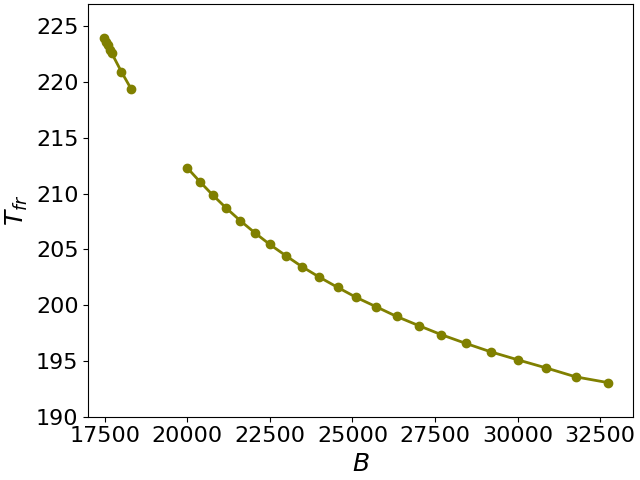}  
    \vspace{-0.2cm}
    \caption{
    Dependence of the period $T_{fr}$ of the loop rotation on the elasto-gravitation number $B$ in the frozen rotating mode reached from a flat inclined circle as the initial configuration. 
    The gap between the dots indicates the range of $B$ corresponding to an irregular mode. 
    }
    \label{fig:period_frozen_rot}
\end{figure}

Frozen rotating mode is observed for relatively elastic loops with values of the elasto-gravitation number $B$ being in the range of $17477 \leq B \leq 32733$. 
In this mode, the loop settles vertically with a constant velocity $v_{z,fr}$ and spins with a constant angular velocity $\omega_{fr}$ around a vertical axis containing the loop centre of mass. 
The angles $\theta$ and $\phi$, plotted in figure~\ref{fig:tilted_angles_vs_time_diiff_modes}, are 
\begin{equation}
  \begin{cases}
    \theta(t) = 90^{\circ}, \\
    \phi(t)   = \omega_{fr} t,
  \end{cases}
  \label{eq:Frozen_angles}
\end{equation}
where the period of rotation $T_{fr}=2\pi/\omega_{fr}$ depends on elasto-gravitation number $B$ as shown in figure~\ref{fig:period_frozen_rot}.
The evolution of the loop's centre of mass is described by the following simple equations,
\begin{equation}
  \begin{cases}
    x_{CM}(t)=x_{c}, \\
    y_{CM}(t)=y_{c}, \\
    z_{CM}(t)=v_{z,fr} t,
  \end{cases}
  \label{eq:frozen_rot_cm}
\end{equation}
where $x_{c}$, $y_{c}$ and $v_{z,fr}$ are constants, and the sedimentation velocity  $v_{z,fr}$ is almost independent on $B$: $1.660 \lesssim \lvert v_{z,fr} \rvert \lesssim 1.664$, as it will be discussed in \textsection\thinspace \ref{sec:di}.

Note that for the selected values of $B$, namely for $B=18622$, $B=18945$, $B=19286$, and $B=19634$, the frozen rotating mode could not be achieved within the monitored simulation time of $88000$.
The final stage of the evolution, illustrated in Supplementary Movies 7 and 8, contains features similar to the irregular mode observed for sedimenting highly elastic fibres~\cite{melikhov_attracting_2024}.

\subsection{Tank-treading, swinging, flapping and irregular modes}\label{sec:othermodes}

In the tank-treading, swinging, flapping, and irregular modes, a significant out-of-plane deformation of the loop shape is observed.
The tank-treading mode was found for $B=34843$. 
The mode is characterised by the loop having a fixed shape that rotates about a vertical axis with a period $T_{rot}=52$ and the constant angular velocity (the azimuthal angle $\phi$ changes linearly with time). 
The centre of mass moves along a helix of a small horizontal circular cross-section with radius $r_{rot} \approx 1$.
The polar angle $\theta = 0 \degree$ but the shape is far from being planar.
In addition, there is also a translation of the beads along the loop with a period $T_{tt}=96$. 
The specifics of the structural configuration and the flow of beads are presented in Movies 9 and 10 of the Supplementary Movies, confirming that this mode is analogous to the tank-treading mode observed in \cite{gruziel-slomka_stokesian_2019}.

The swinging mode, analogous to the swinging mode reported in \cite{gruziel-slomka_stokesian_2019}, is observed for two values of the elasto-gravitation number: $B=35993$ and $37244$, with the periods $T_{swing}=303$ and $308$, respectively. 
It involves significant time-dependent periodic deformations of the loop shape, as shown in Movie 11 of the Supplementary Movies. 
At each time instant, the shape is symmetric with respect to a vertical plane ($xz$ plane in Movie 11). 
Therefore, the motion of the centre of mass is restricted to this plane.
Two cases of the swinging mode exhibit slight differences in the one-dimensional horizontal drift of the centre of mass.
For $B=35993$, the centre of mass of the loop exhibits symmetric oscillations with the amplitude $7d$ about a fixed position, resulting in no horizontal displacement after a single swinging cycle.
In contrast, for $B=37244$, the forward and backward motions of the centre of mass of the loop are not symmetric, resulting in a horizontal displacement of $2 d$ after each swinging cycle.

The flapping mode is observed for $B=40000$ with the period of $T_{flap}=157$. 
The loop shape undergoes considerable time-dependent periodic deformations, as shown in Movie 12 of the Supplementary Movies. 
The configuration changes from an almost flat and almost horizontal shape to a severely bent 3D shape. 
In contrast to the swinging mode, all subsequent configurations possess two vertical planes of symmetry perpendicular to each other. 
Therefore, the centre of mass does not move horizontally.
This mode is also analogous to the flapping mode reported in \cite{gruziel-slomka_stokesian_2019}.

The irregular mode is observed for $B=33746$, and for several other values of $B$, as discussed in \textsection\thinspace \ref{sec:frozenrotatmode}.
The mode is characterised by irregular rotations and oscillations of the centre of mass trajectory and considerable irregular deformations of the loop shape with time. 
This mode is analogous to the irregular mode observed in the case of sedimenting fibres reported in \cite{melikhov_attracting_2024}.

\subsection{Diagram of the modes}

Different attracting dynamical modes, reached from the initially inclined circular configuration, for different ranges of the elasto-gravitation number $B$, are shown as a phase diagram in figure~\ref{fig:phase-diag}. 
The evolution of shapes in these modes is illustrated in the Supplementary Movies.

\begin{figure}
    \centering
    \includegraphics[width=0.91\textwidth]{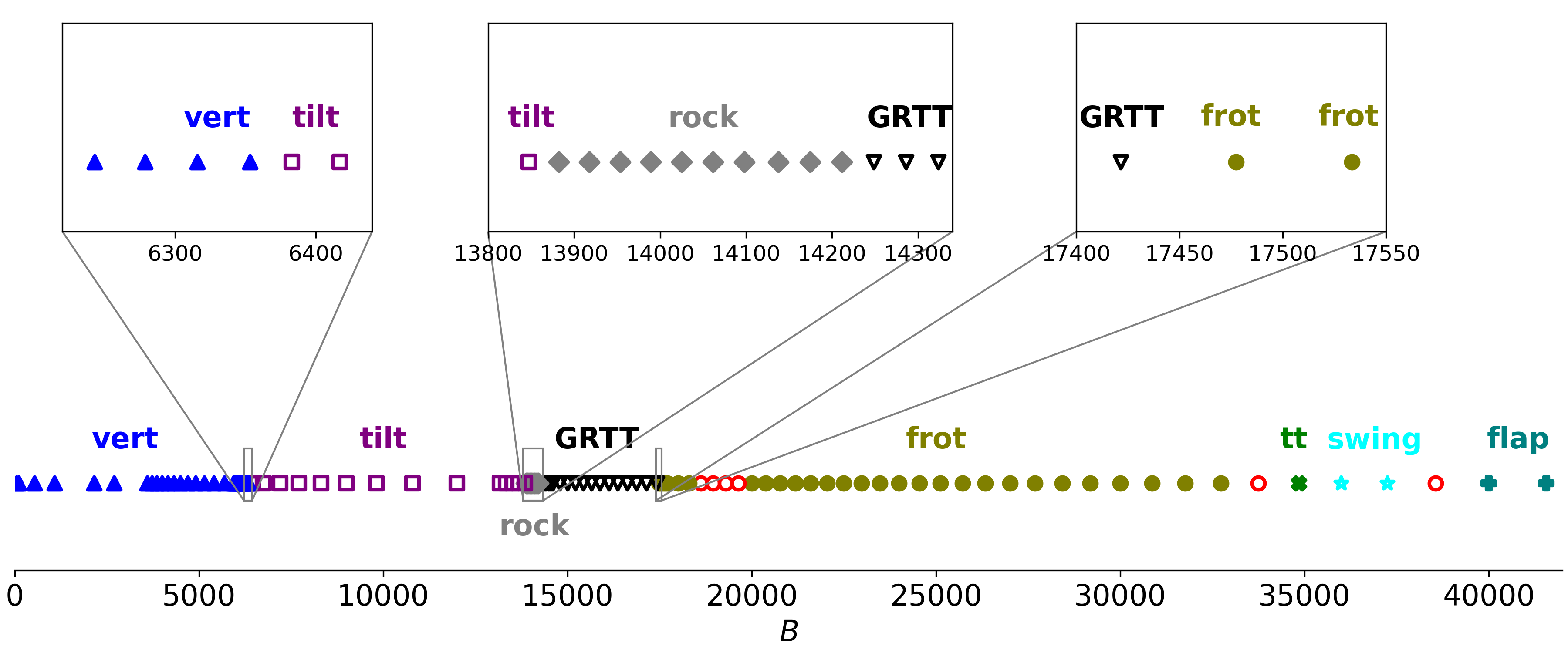}
    \vspace{-0.2cm}
    \caption{
    Dependence of attracting dynamical modes on elasto-gravitation number $B$. 
    Initially, the loop is planar, circular, and inclined, typically at 
    $\theta\!=\!16\degree$. 
    Critical values of $B$ separating distinct modes are: $6366$ (vertical-tilted), $13877$ (tilted-rocking), and $14218$ (rocking-GRTT). 
    The empty circles correspond to an irregular mode. 
    }
    \label{fig:phase-diag}
\end{figure}

\section{The frozen rotating mode reached from a non-planar initial configuration}\label{sec:bi}

\begin{figure}
    \centering
    \includegraphics[width=8cm]{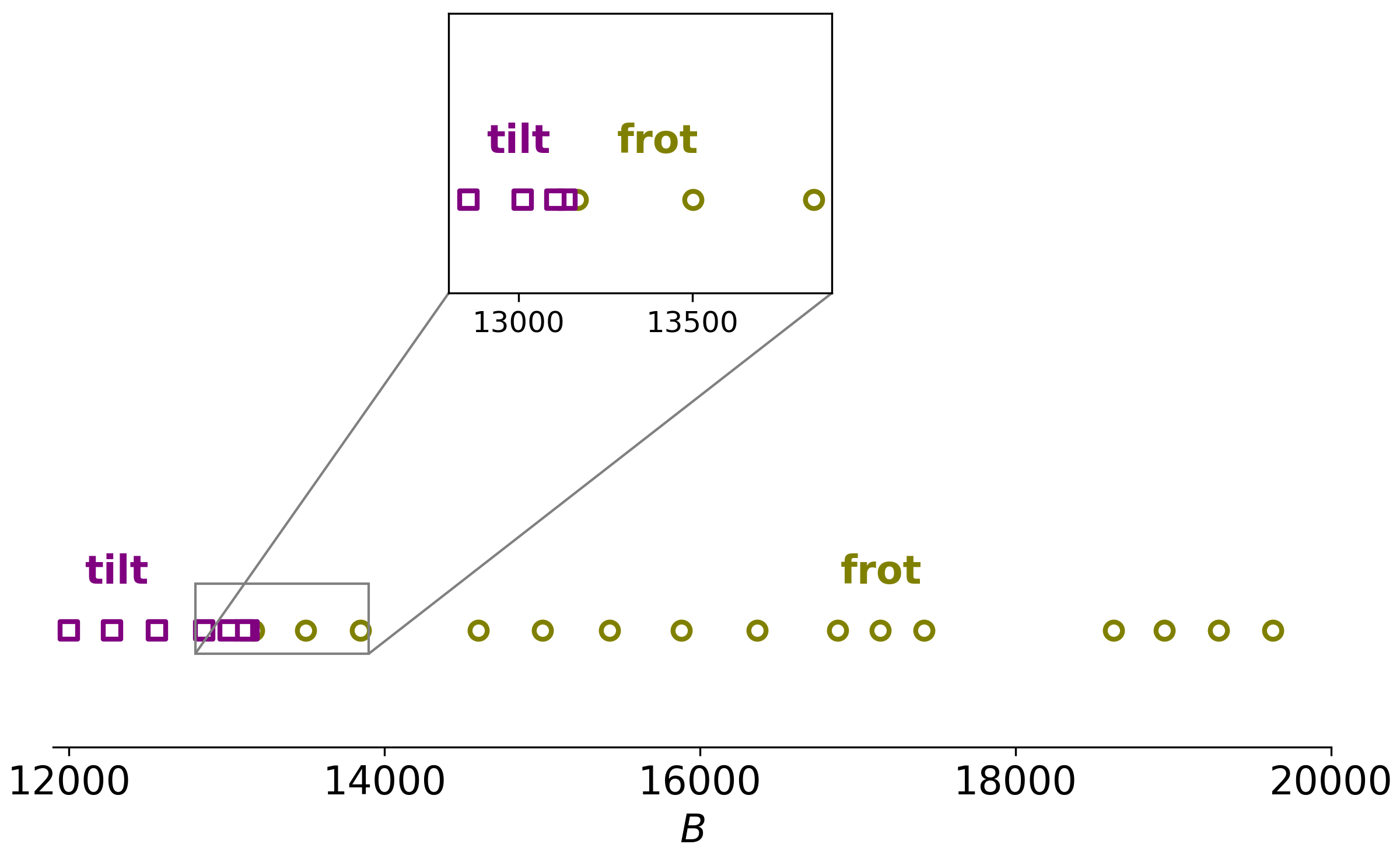}
    \vspace{-0.2cm}
    \caption{
    Dependence of attracting dynamical modes on elasto-gravitation number $B$ in the case of the non-planar initial configuration. 
    The critical value of $B$ separating the tilted and frozen rotating modes is $13167$.
    }
    \label{fig:phase-diag-curved-init-cond}
\end{figure}

It is known that different attracting dynamical modes can coexist, reached from different initial configurations, as shown in figure~20 in \cite{gruziel-slomka_stokesian_2019}. 
Of special interest is the frozen rotating mode, with a fixed shape. 
Therefore, in this work, we check if the %
frozen rotating mode can exist for a wider range of $B$ if a different initial configuration is chosen. %
As the non-planar initial configuration, we took the final configuration in sedimentation of the loop with $B=17477$, i.e., the loop that ended up in the frozen rotating mode. 
We performed numerical simulations for
$12000 \leq B \leq 17422$  and %
$18622 \leq B \leq 19634$. The last range %
corresponds to the values of $B$ that result in irregular modes if a flat inclined circle as the initial configuration was selected. For the non-planar initial configuration, the loop ends up in the frozen rotating mode with a shape slightly different than the initial one. 

\begin{figure} 
    \centering
    \includegraphics[width=0.46\textwidth]{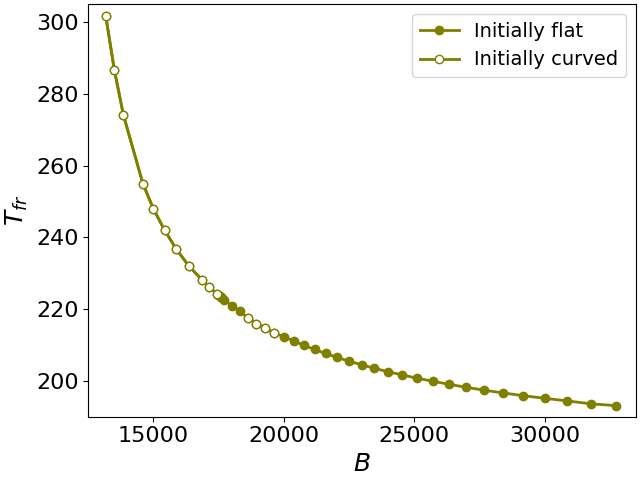}
    \vspace{-0.2cm}
    \caption{
    Dependence of the period $T_{fr}$ of the loop rotation on the elasto-gravitation number $B$ in the frozen rotating mode approached from two different initial conditions: a flat inclined circle, or a non-planar loop, as explained in the text. 
    }
    \label{fig:period_frozen_rot_all}
\end{figure}

For $12000 \leq B \leq 17422$, if a flat inclined initial configuration is chosen, the loop ends up in the tilted, rocking, or GRTT modes, as shown in figure~\ref{fig:phase-diag}. 
However, if the non-planar initial configuration is selected, the tilted mode is approached in a narrower range of values, $12000 \leq B \leq 13138$, and for larger values, $13169 \le B \le 17422$, the frozen rotating mode appears, as shown in  figure~\ref{fig:phase-diag-curved-init-cond}. 
The rotation period in the frozen rotating mode, reached from both flat and non-planar initial configurations, is plotted in figure~\ref{fig:period_frozen_rot_all}, now in a wider range than in figure~\ref{fig:period_frozen_rot}. 

\begin{figure}
    \centering
    \includegraphics[width=0.91\textwidth]{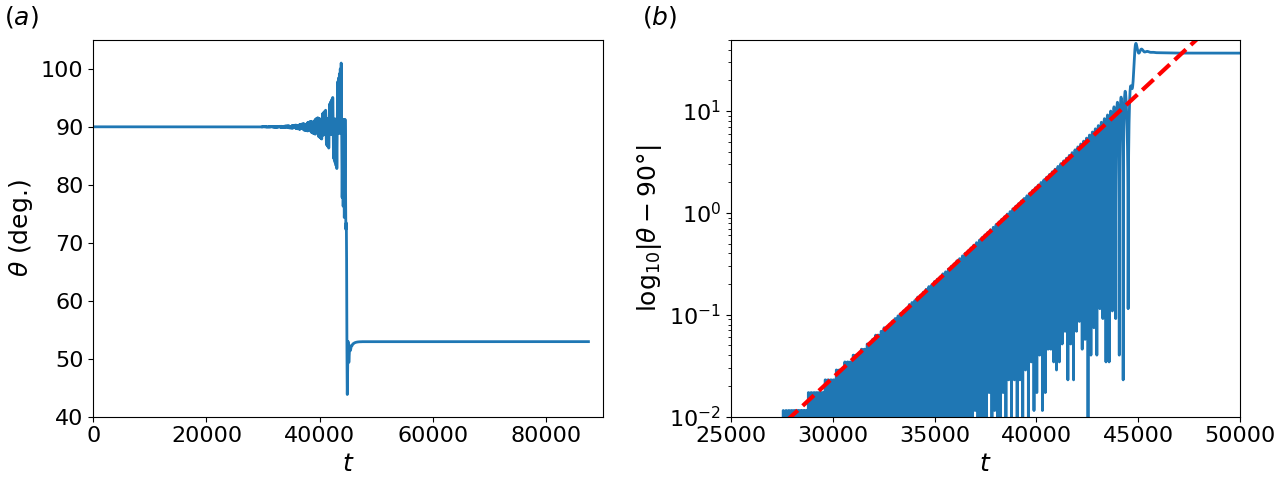}
    \vspace{-0.2cm}
    \caption{%
    Dependence of $\theta$ on time for the non-planar initial configuration and an exemplary value $B=12558$. 
    (\textit{a})~%
    The frozen rotating mode is quasi-stable until $t \approx 30000$ followed by a destabilisation till $t \approx 45000$ when the tilted mode is formed. 
    (\textit{b})~The amplitude $A_o$ of the oscillations of $\theta$ in the destabilization stage is well-approximated by \eqref{eq:theta-90_frozen} with $\lambda = 4.28 \cdot 10^{-4}$, shown as the straight dashed line. 
    }
    \label{fig:frozen_rot_theta_vs_time}
\end{figure}

To estimate the critical value $B_{crit}$ of the transition between the tilted and frozen rotating modes, the evolution $\theta(t)$ is studied for the cases in which the frozen rotating mode destabilizes if starting from the non-planar initial configuration. 
An example of this analysis is shown in figure~\ref{fig:frozen_rot_theta_vs_time} for $B=12558$. 
It is seen that $\theta \approx 90\degree$ for a while, but after the transition period, the loop continues to sediment in the tilted mode.

\begin{figure}
    \centering
    \includegraphics[width=0.46\textwidth]{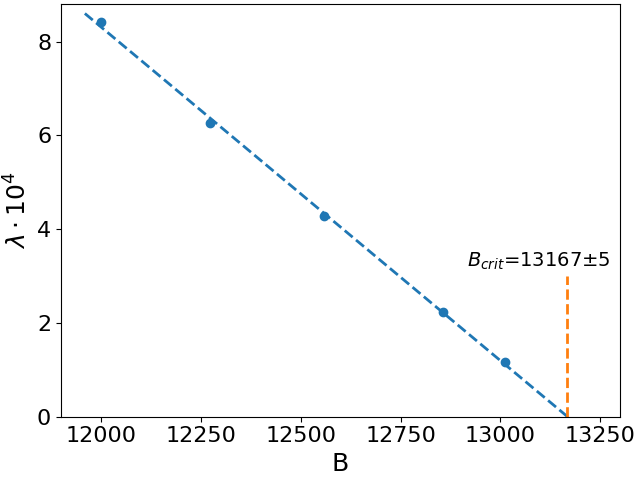}
    \vspace{-0.2cm}
    \caption{
    The growth rate $\lambda$ is a linear function of $B$, enabling estimation of $B_{crit}$=13167. 
    }
    \label{fig:frozen_rot_theta_vs_time_bis}
\end{figure}

The initial stage of destabilization can be analysed via the dependence of $\theta$ on time. 
The time-dependent amplitude $A_o(t)$ of the oscillations of $\theta$ around $90 \degree$ 
can be approximated to increase exponentially:
\begin{equation}
  A_o(t) \propto e^{\lambda t}
  \label{eq:theta-90_frozen}
\end{equation}
in the range $12000 \le B \le 13012$.
The dependence of $\lambda>0$ on $B$ is linear, allowing identification of the value of $B_{crit} \approx 13167$ at which frozen rotating becomes unstable, as illustrated in figure~\ref{fig:frozen_rot_theta_vs_time_bis}.

\section{Centre-of-mass velocity and characteristic time scales for the attracting modes}
\label{sec:time}

\begin{figure}
    \centering
    \includegraphics[width=0.91\textwidth]{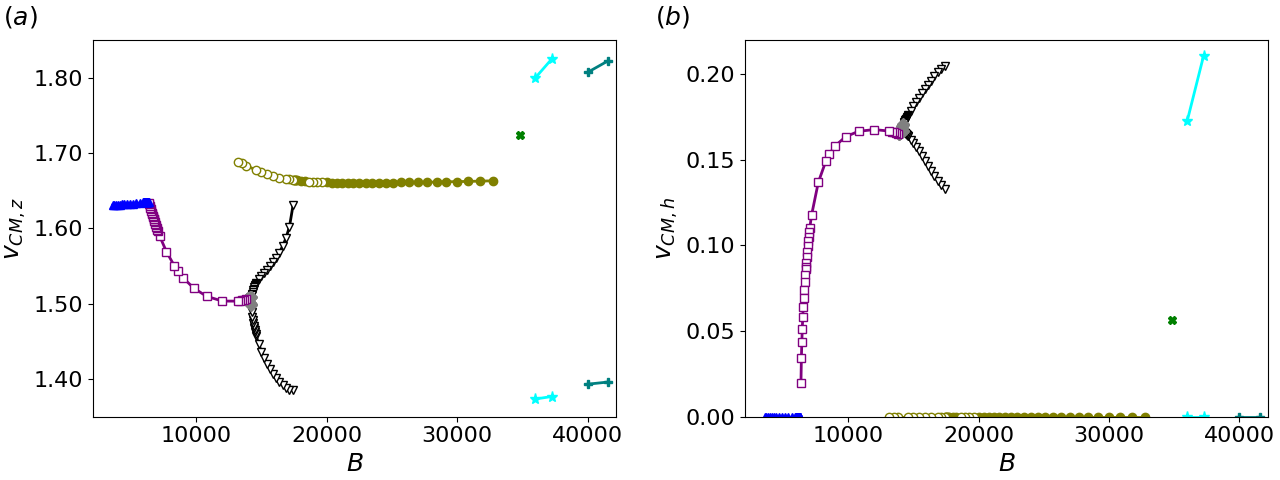}
    \vspace{-0.2cm}
    \caption{
    (\textit{a})~The sedimentation velocity $v_{CM,z}> 0$ and 
    (\textit{b})~the lateral velocity $v_{CM,h}\ge 0$ of the loop centre-of-mass as a function of elasto-gravitation number $B$.
    For the rocking, GRTT, swinging and flapping modes, the maximum and minimum of both velocities are shown (the average of $v_{CM,h}$ for the GRTT mode is shown in figure~\ref{fig:radius_grtt_cm_gyration}(\textit{b})). 
    In the vertical, tilted, frozen rotating, and tank-treading modes, both velocities do not depend on time. 
    The colours and symbols correspond to the modes as presented in figures~\ref{fig:phase-diag} and \ref{fig:phase-diag-curved-init-cond} for the planar and non-planar initial configurations, respectively. 
    }
    \label{fig:sed_veloc}
\end{figure}

In figure~\ref{fig:sed_veloc}, we compare the characteristic centre-of-mass velocity of the elastic loop made of 36 spherical beads in different attracting modes.  
In panels (a) and (b), we show separately the absolute values of the vertical and horizontal components of the centre-of-mass velocity, $v_{CM,z}$ (in the following called the sedimentation velocity) and $v_{CM,h}$ (called the lateral velocity).
The chosen normalization, given in \eqref{normal}, is such that the sedimentation velocity of a single bead is equal to $1/3$.

The main observation is that the sedimentation velocity is typically at least an order of magnitude larger than the lateral one. 
There is no horizontal drift in the vertical, frozen rotating, and flapping modes. 
The lateral velocity $v_{CM,h}$ is constant in time for the tank-treading and tilted modes. 
In the tilted mode, the horizontal drift tends to vanish when $B$ decreases to the critical value corresponding to the transition between the tilted and vertical modes. 
In the tilted mode, the horizontal component of the CM velocity has a fixed direction, and in the tank-treading mode, the horizontal component of the CM velocity corresponds to the motion along a circle. 

In the rocking mode, the horizontal component of the CM velocity has a fixed direction plus oscillations along this direction and oscillations perpendicular to it. 
In the swinging mode, the horizontal component of the CM velocity has a fixed direction plus oscillations along this direction. 
In the GRTT mode, the horizontal component of the CM velocity corresponds to the motion along a circle plus periodic oscillations along the circle and perpendicular to it. 
The magnitude of the oscillations about the mean values in the rocking and GRTT modes increases with $B$, seen as the increase in the difference between the maximum and minimum values of the sedimentation and lateral velocities.

The dependence of the sedimentation and lateral velocities on $B$ is non-monotonic and very complex. 
The loop shape \& orientation, and their variation with time, are essential for the resulting value and direction of the centre-of-mass velocity. 
There is no systematic dependence of the lateral velocity on $B$. 
The sedimentation velocity of the loop with a fixed shape in the tank-treading mode is larger than in the frozen rotating mode, in the frozen rotating mode -- larger than in the vertical mode, and in the vertical mode -- larger than in the tilted mode. 
The last inequality is in agreement with the behaviour of an inclined rigid circle that sediments faster than a vertical circle of the same diameter.

\begin{figure}
    \centering
    \includegraphics[width=0.91\textwidth]{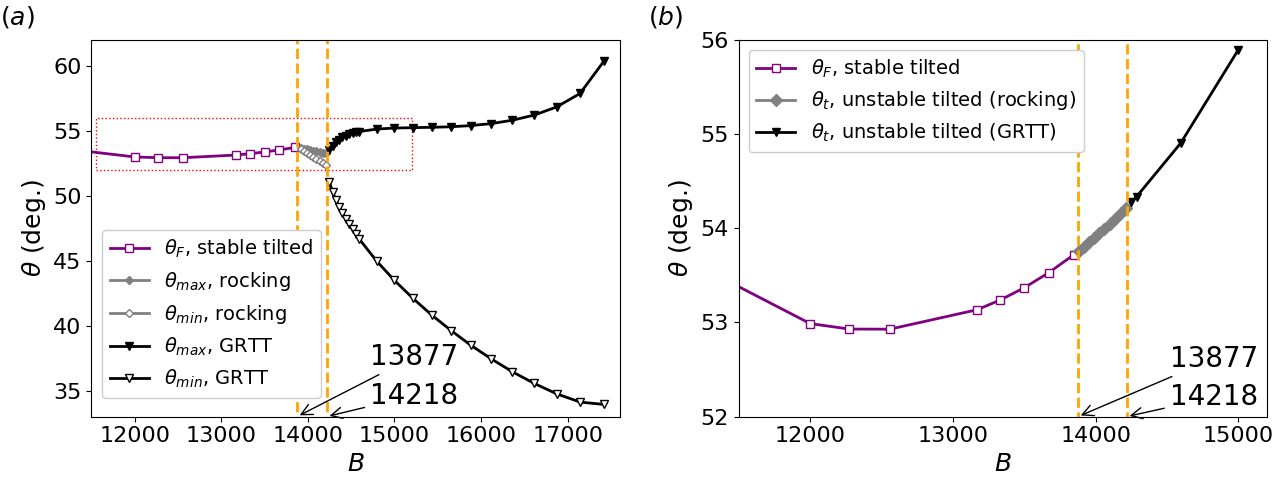}
    \vspace{-0.2cm}
    \caption{
    The polar angle $\theta$ of the loop as a function of the elasto-gravitation number $B$.
    (\textit{a})~The final $\theta_{F}$ for the tilted mode
    and maximum and minimum values of $\theta$ during the rocking and GRTT modes, $\theta_{max} = \theta_{0} + \theta_{2}$ and $\theta_{min} = \theta_{0} - \theta_{2}$, see figures~\ref{fig:tchar_tilted_vertical}(\textit{a}) and \ref{fig:Rocking_Bcrit}(\textit{a}).
    (\textit{b})~The polar angle for the stable tilted mode($\theta_{F}, \;B\le 13877$) and for unstable tilted configuration ($\theta_{t}, \;12877\le B\le 15000$) before the rocking and GRTT modes are established. 
    The dashed rectangle in (\textit{a}) corresponds to the axes limits used in (\textit{b}).
    }
    \label{fig:Theta_vs_B_all_modes}
\end{figure}

For a single mode with the shape deformation, variations of the sedimentation velocity can be as large as 30\%. 
In general, it might be expected that almost horizontal loop (with a smaller value of $\theta$) sediments slower than almost inclined loop (with a greater value of $\theta$), and therefore, that the lower and upper branches in figure~\ref{fig:sed_veloc} (the minima and maxima of $v_{CM,z}(t)$ over time) correspond to the lower and upper branches in figure~\ref{fig:Theta_vs_B_all_modes}(a) (the minima and maxima of $\theta(t)$), respectively. 
Indeed, for the rocking mode, the time positions of $\theta_{max}$ and $\theta_{min}$ coincide with the time positions of the maximum and minimum values of $v_{CM,z}$. 
For the GRTT and swinging modes, the values of the time at which $\theta_{max}$ and $\theta_{min}$ are reached are close to the values of the time corresponding to the maximum and minimum values of $v_{CM,z}$. 
The reason for a small difference could be assigned to non-planar and non-circular shapes. 
For the flapping mode, the sedimentation velocity is the smallest when the loop is almost horizontal.

We now compare the characteristic timescales of the attracting modes, ordering them from the shortest to the longest. 
For all the modes, the smallest is the sedimentation time scale, of the order of 0.5-0.7. 
The time scale of the horizontal drift is $\gtrsim 5$. 
The time scales of the rotation and tank-treading in the tank-treading mode are 50 and 100, of the flapping in the flapping mode -- 150, of the rotation in the frozen rotating mode -- 200-300, of the rocking in the rocking and GRTT modes and swinging in the swinging mode -- around 300. 
Much longer are the gyration (1000-50000) and tank-treading (4000-100000) time scales in the GRTT mode. 
In general, the attracting modes are reached after a long time.

\section{Discussion and conclusions}
\label{sec:di}

In our numerical findings, all the stationary configurations of the highly elastic loops are non-planar except the vertical one. 
This result differs from the known fact that all configurations of rigid circles are stationary -- they translate vertically and horizontally with no change in their inclination angle, and are neutrally stable, as the sedimenting rods \cite{taylor_1967}. 

Now we will argue that there are no planar circular stationary configurations for an elastic loop made of $N$ beads, with the inclination angle $\theta$ other than $0\degree$ (horizontal). %
Assume that there is one, which means that the velocity vector $\bm{v}_i$ of each bead $i=1,...,N$ is the same. 
The vectors $\bm{v}_i$ are sums of two contributions: $\bm{v}_i^{\perp}$, 
caused by the forces (acting on all the beads $j=1,...,N$) {\it perpendicular} to the circle, and $\bm{v}_i^{\parallel}$  
caused by the forces (acting on all the beads $j=1,...,N$) {\it parallel} to the circle. %
The forces perpendicular to the circle are only the perpendicular components of gravity, while the parallel ones are the parallel components of gravity plus all the elastic forces. 
Owing to symmetry, $\bm{v}_i^{\perp}$ is perpendicular, and $\bm{v}_i^{\parallel}$ is parallel  to the circle; moreover, $\bm{v}_i^{\perp}$ is the same for all the beads $i$. 

To require that the vector $\bm{v}_i^{\parallel}$ is also the same for all the beads, we need to select the circle in the elastic equilibrium (with vanishing elastic forces). 
In particular, for the horizontal circle (i.e., for $\theta=0$) in the elastic equilibrium, no horizontal forces are acting on each bead $i$; therefore, this configuration is stationary. 
In general, $\theta \ne 0$, and the only parallel force acting on a bead $i$ is the parallel component of gravity, $F=G/N \sin \theta \ne 0$.  
We consider $N=36$ for simplicity and assume that beads 1 and 19 are at the bottom and at the top, respectively, and the line of centres of beads 10 and 28 is horizontal. 
Numerical evaluation of $\bm{v}_i^{\parallel}$ demonstrates that $|\bm{v}_{10}^{\parallel}|>|\bm{v}_1^{\parallel}|$ by around 6\%, in a qualitative agreement with analytical predictions based on the point-force model (the details are given in Appendix~\ref{no-circle}). Therefore, a circle is not a stationary configuration, except if it is oriented horizontally. 

In the vertical attracting mode, a stationary non-circular planar (vertical) configuration is reached, shown in figure~\ref{fig:vertical_shape} in Appendix \ref{no-circle}, and also in figure 2 in \cite{gruziel-slomka_stokesian_2019}. However, this configuration is not stationary if tilted -- our numerical computations show that $\bm{v}_i^{\perp}$ is not the same for all the beads $i$ (see Appendix~\ref{no-circle}). 
These observations seem to indicate the non-existence of a planar stationary configuration of an elastic loop, except if oriented vertically or horizontally.  
Our numerical analysis demonstrates that the stationary horizontal circular configuration is unstable, as shown in Appendix~\ref{no-circle}. 

We find also other unstable stationary configurations: unstable frozen rotating and unstable tilted stationary configurations, as shown in figures~\ref{fig:frozen_rot_theta_vs_time} and~\ref{fig:Rocking_Bcrit}, respectively.  
The averaged inclination angle $\theta_t$ of the unstable tilted mode is shown in figure~\ref{fig:Theta_vs_B_all_modes}(b) as a function of $B \ge 13877$. 
It is clear that in this range of $B$, the function $\theta_t(B)$ for the unstable stationary configuration extends smoothly  the function $\theta_F(B)$ corresponding to the stable stationary configuration with $B\le 13877$. 
The existence of stable and unstable stationary tilted and frozen rotating configurations is important for the overall structure of invariant manifolds of the dynamics, and their stability, as described, e.g., in \cite{Barenblatt1996}. 
In future studies of the invariant manifolds and bifurcations of the dynamics of highly elastic loops (or fibres with open ends), the methods from \cite{fox2024data} might be useful. 

The attracting dynamical modes of highly elastic loops are similar to the attracting modes of highly elastic fibres with open ends, described in \cite{melikhov_attracting_2024}. 
In particular, the analogues of the rocking and GRTT modes for loops are, respectively, the crawling and rotation-crawling modes for the fibres.
In general, the attracting modes are approached after a long time.

In conclusion, the main results of this paper are as follows. 
Different attracting dynamical modes of a highly elastic loop are observed for different values of the elasto-gravitation number $B$, starting from planar and non-planar initial conditions. 
This work extends the study performed by \cite{gruziel-slomka_stokesian_2019} as outlined in Appendix~\ref{cGS}. 
Two new modes of the dynamics are described: rocking and GRTT, and it is shown that (as the tilted mode) they can coexist with the frozen rotating mode for certain values of $B$. 
Critical values of $B$ corresponding to transitions between all the attracting modes are determined. 
Beyond the transition, some of the modes become unstable (but still can be observed), while others cease to exist. 
For each attracting mode, characteristic time scales are evaluated. 
Sedimentation and lateral translational velocities, and the angular velocity are calculated and shown to vary depending on the mode and the value of the elasto-gravitation number. 

The differences between the velocities determined here might be used in a centrifuge to sort the loops with different densities, or bending stiffness. 
A similar sorting of elastic fibres with open ends can be expected, based on the results shown in figure 5(d) in \cite{melikhov_attracting_2024}. 
The dynamics of highly elastic loops and fibres may be useful to explain the motion and shape deformation of more complex highly elastic objects, e.g., disks or sheets \cite{miara2022sedimentation,yu2024free}.

\section*{Acknowledgements} \label{ackn}
\textit{
    This work was supported in part by the National Science Centre under grant UMO-2021/41/B/ST8/04474. 
    We thank Magdalena Gruziel-S\l omka, Piotr Szymczak, Maciej Lisicki and Radost Waszkiewicz for helpful discussions. 
}

\section*{Declaration of interests}
\textit{
    The authors report no conflict of interest.
}

\section*{Data availability statement}
\textit{
    The data that support the findings of this study are openly available in RepOD - Repository for Open Data at \url{https://doi.org/10.18150/IOI0RG}. 
}

\appendix

\section{Description of the movies}
\label{app:movies}
Note that the diameter of all beads in all movies is reduced to facilitate the visualisation of motion details. 
For clarity, different colours are used to distinguish individual beads. 
We use the reference frame translating with the centre-of-mass, and rotating with the average angular velocity, denoting the coordinates as  $\tilde{x}$, $\tilde{y}$ and $\tilde{z}$.
Gravity acts along the negative $\tilde{z}$-direction.
For clarity, the tick labels are specified only on the $\tilde{z}$-axis. 
The distance between the major ticks on the axes $\tilde{x}$, $\tilde{y}$ and $\tilde{z}$ is the same.

Movie 1. Tilted mode:
The emergence of a tilted mode for $B=13501$, starting from the loop that is initially planar, circular, and inclined at $\theta = 16 \degree$. 
The centre-of-mass motion is subtracted.
The tilted mode is achieved at $t \approx 3500$.

Movie 2. Tilted mode:
Structural details of the tilted shape for $B=13501$.

Movie 3. Rocking mode:
Visualisation of the rocking mode for $B=14061$. 
The motion with a period $T_{r} = 297$ is seen. 
The centre-of-mass motion is subtracted.
Two complete cycles of the rocking motion are presented.

Movie 4. Rocking mode:
Structural details of the rocking shape for $B=14061$ at a selected time instant.

Movie 5. GRTT mode:
Visualisation of the GRTT mode for $B=15428$ (view 1).
The rocking part of the motion with a period $T_{r} = 276$ is seen.
Three complete cycles of the rocking motion are presented.
The centre-of-mass motion is subtracted, together with the averaged rotational motion with a period $T_{g} = 4797$.

Movie 6. GRTT mode:
Visualisation of the GRTT mode for $B=15428$ (view 2).
The time range is identical to that in view 1, but the elevation and azimuth angles for viewing are adjusted for a different perspective.
Note that the initial portion of the tank-treading motion cycle with $T_{tt} = 14352$ is visible, with the black bead clearly moving along the loop by the end of the third rocking cycle. 
The centre-of-mass motion is subtracted, together with the averaged rotational motion with a period $T_{g} = 4797$.

Movie 7. Frozen rotating mode:
Visualisation of the frozen rotating mode for $B=22497$. 
The centre-of-mass motion is subtracted. 
The motion with a period $T_{fr} = 205$ is seen.
Two complete cycles of the motion are presented.

Movie 8. Frozen rotating mode:
Visualisation of the frozen rotating mode for $B=22497$ showing that the shape of the loop does not change during the frozen rotating mode.
The centre-of-mass motion is subtracted, together with the rotational motion with a period $T_{fr} = 205$.

Movie 9. Tank-treading mode:
Visualisation of the tank-treading mode for $B=34843$.
The motion with a period $T_{tt} = 96$ is seen.
Two complete cycles of the motion are presented.
The centre-of-mass motion is subtracted, together with the averaged rotational motion with a period $T_{g} = 52$.

Movie 10. Tank-treading mode:
Structural details of the tank-treading shape for $B=34843$ at a selected time instant.

Movie 11.%
Visualisation of the swinging mode for $B=35993$. 
The centre-of-mass motion is subtracted. 
The motion with a period $T_{swing} = 303$ is seen.
Two complete cycles of the swinging motion are presented from two perspectives, separated by structural details of shape at a selected time instant. 
Note that the loop shape is symmetric with respect to reflection in the $\tilde{x}\tilde{z}$-plane.

Movie 12. %
Visualisation of the flapping mode for $B=40000$. 
The centre-of-mass motion is subtracted. 
The motion with a period $T_{flap} = 157$ is seen.
Two complete cycles of the flapping motion are presented from two perspectives, separated by structural details of shape at a selected time instant.
Note that the loop shape has two planes of symmetry:
$\tilde{x}\tilde{z}$ and $\tilde{y}\tilde{z}$.

\section{Fitting expressions}
\label{app:fits}

For the rocking and GRTT modes, we perform the fast Fourier transform (FFT) of the numerical data for the time-dependent orientation angles $\theta$ and $\phi$ and the centre-of-mass position $(x_{CM},\,y_{CM},\,z_{CM})$ and we determine the leading frequencies. 
Then, we use these values to construct and plot the fitting expressions. 

\begin{figure}
    \centering
    \includegraphics[width=0.85\textwidth]{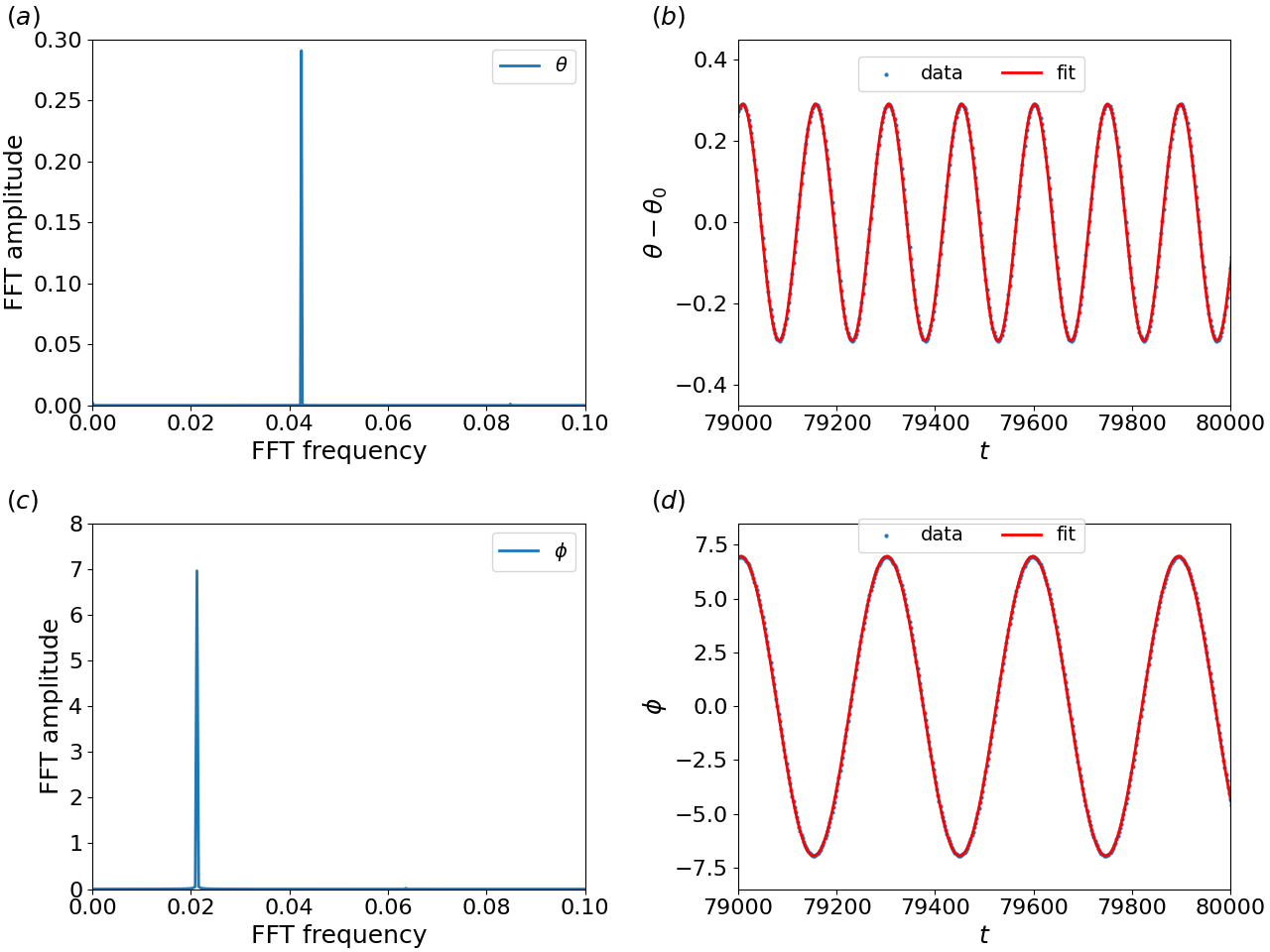}
    \vspace{-0.2cm}
    \caption{
    Fitting the time dependence given by \eqref{eq:rocking_angles} to the numerical data for the orientation angles $\theta$ and $\phi$ in the rocking mode with $B=14098$. 
    (\textit{a})~and (\textit{c})~The FFT analysis of $\theta$ and $\phi$. 
    (\textit{b})~The fit of $\theta-\theta_0$ is performed with one dominant FFT frequency $2 \omega_r$. 
    (\textit{d})~The fit of $\phi$ is performed with one dominant FFT frequency $\omega_r$. 
    }
    \label{fig:fit_rocking_angles}
\end{figure}

\begin{figure}
    \centering
    \includegraphics[width=0.85\textwidth]{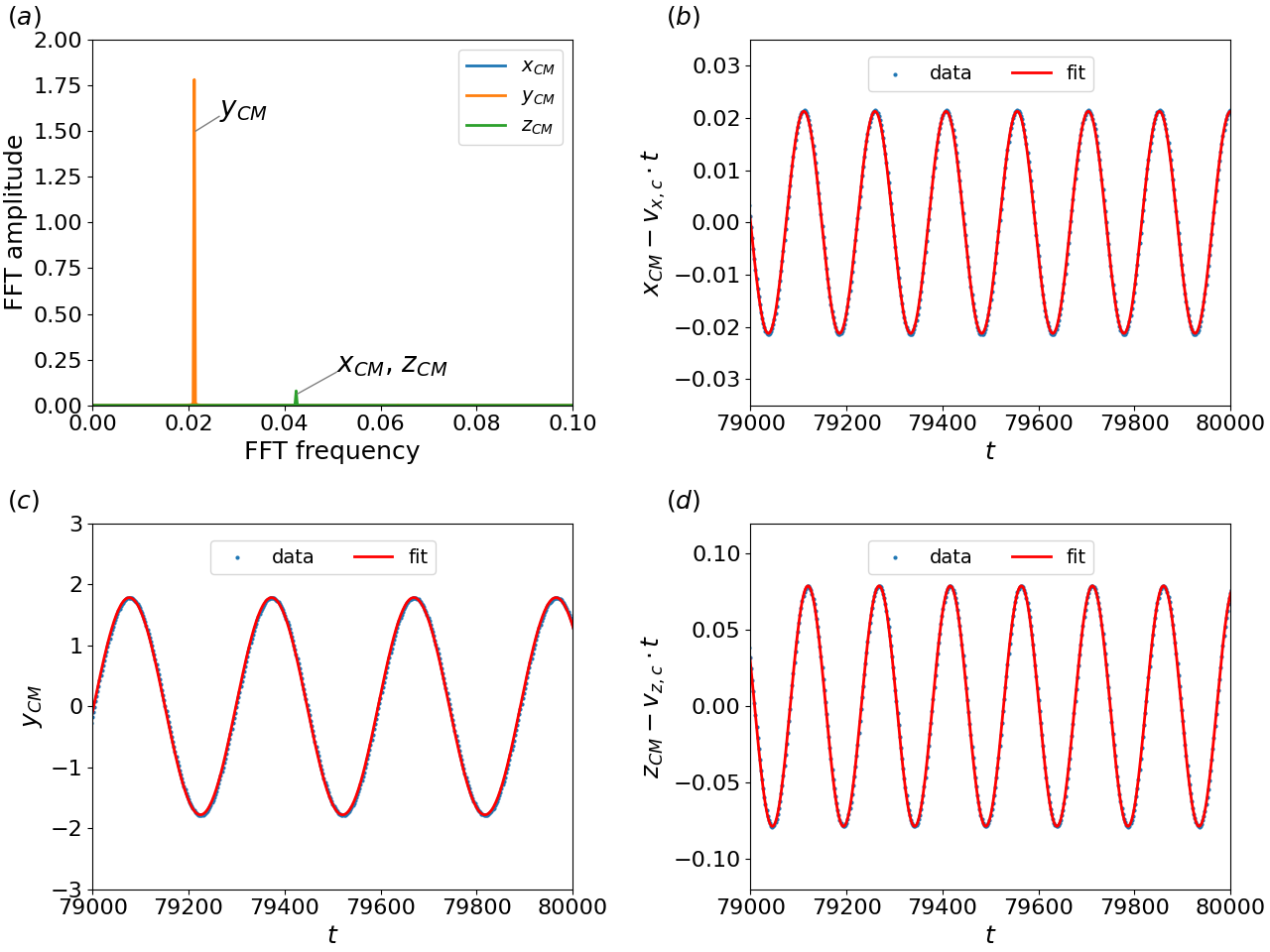}
    \vspace{-0.2cm}
    \caption{
    The centre-of-mass movement for the rocking mode with $B=14098$. 
    (\textit{a})~The FFT of $x_{CM}$, $y_{CM}$ and $z_{CM}$. 
    (\textit{b})-(\textit{d})~The fits of quasi-periodic $x_{CM}$, $y_{CM}$ and $z_{CM}$ were performed with one dominant FFT frequency only: $\omega_{r}$ for $y_{CM}$, and $2 \omega_{r}$ for $x_{CM}$ and $z_{CM}$.
    The amplitudes are $0.0213$, $1.781$, and $0.079$ for $x_{CM}$, $y_{CM}$ and $z_{CM}$, respectively.
    See \eqref{eq:rocking_cm} for the fitting functions. 
    }
    \label{fig:fit_rocking_cm}
\end{figure}

The results of the FFT analysis of the data and the fitting expressions for the rocking mode with $B=14098$ are shown in figures~\ref{fig:fit_rocking_angles}-\ref{fig:fit_rocking_cm}. 
The parameters are 
$\theta_0 = 53 \degree$, 
$\omega_r = 2.12 \cdot 10^{-2}$, %
$v_{x,c} = 0.166$, 
and 
$v_{z,c} = 1.505$. 

\begin{figure}
    \centering
    \includegraphics[width=0.85\textwidth]{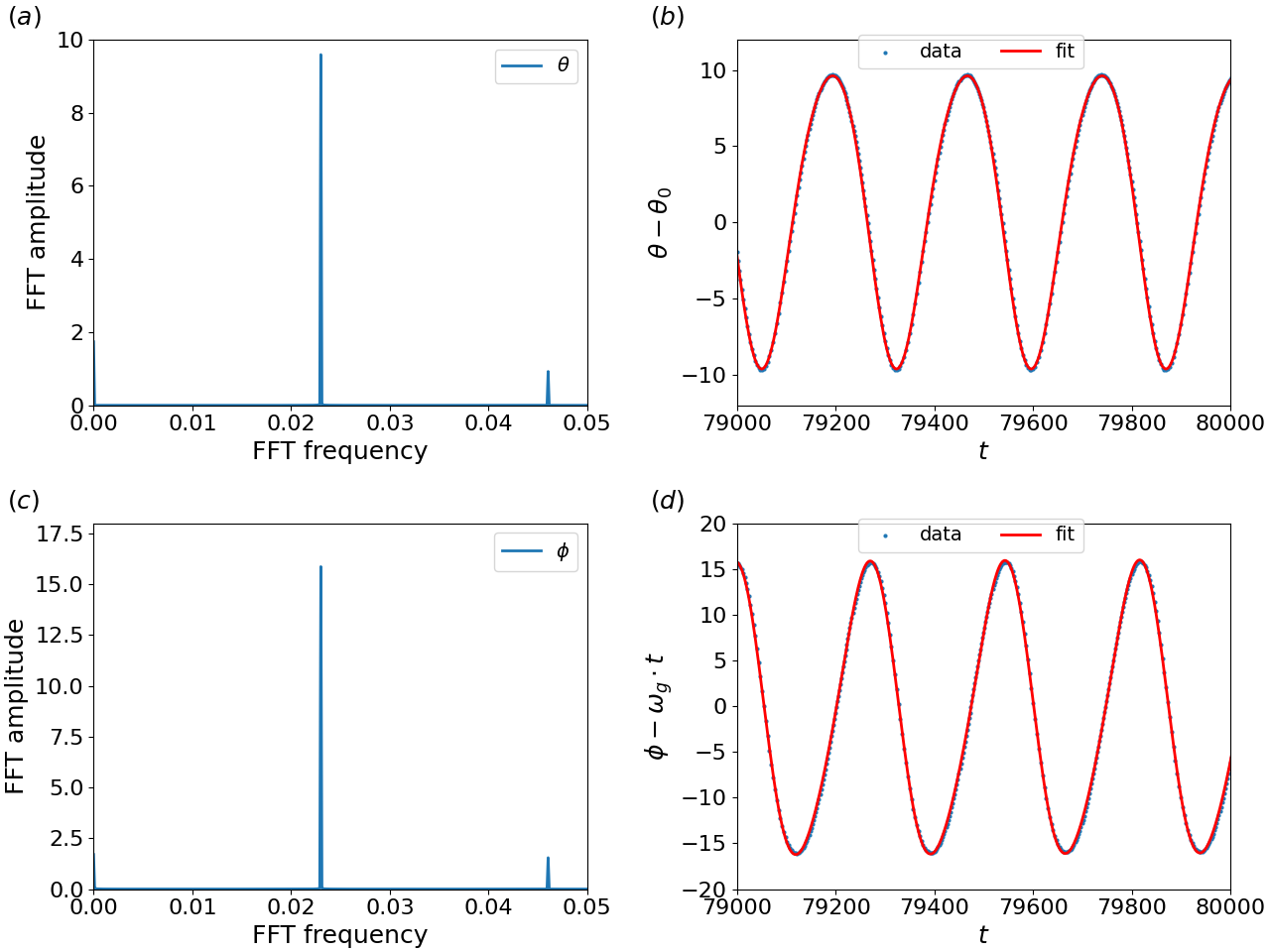}
    \vspace{-0.2cm}
    \caption{
    Fitting the time dependence given by \eqref{eq:GRTT_angles} to the numerical data for the orientation angles $\theta$ and $\phi$ in the GRTT mode with $B=16362$. 
    (\textit{a})~and (\textit{c})~The FFT analysis of $\theta$ and $\phi$. 
    (\textit{b})~and (\textit{d})~The fits of $\theta-\theta_0$ and $\phi -\omega_g t$, performed with two dominant FFT frequencies, $\omega_r$ and $2\omega_r$. 
    }
    \label{fig:fit_grtt_angles}
\end{figure}

\begin{figure}
    \centering
    \includegraphics[width=0.85\textwidth]{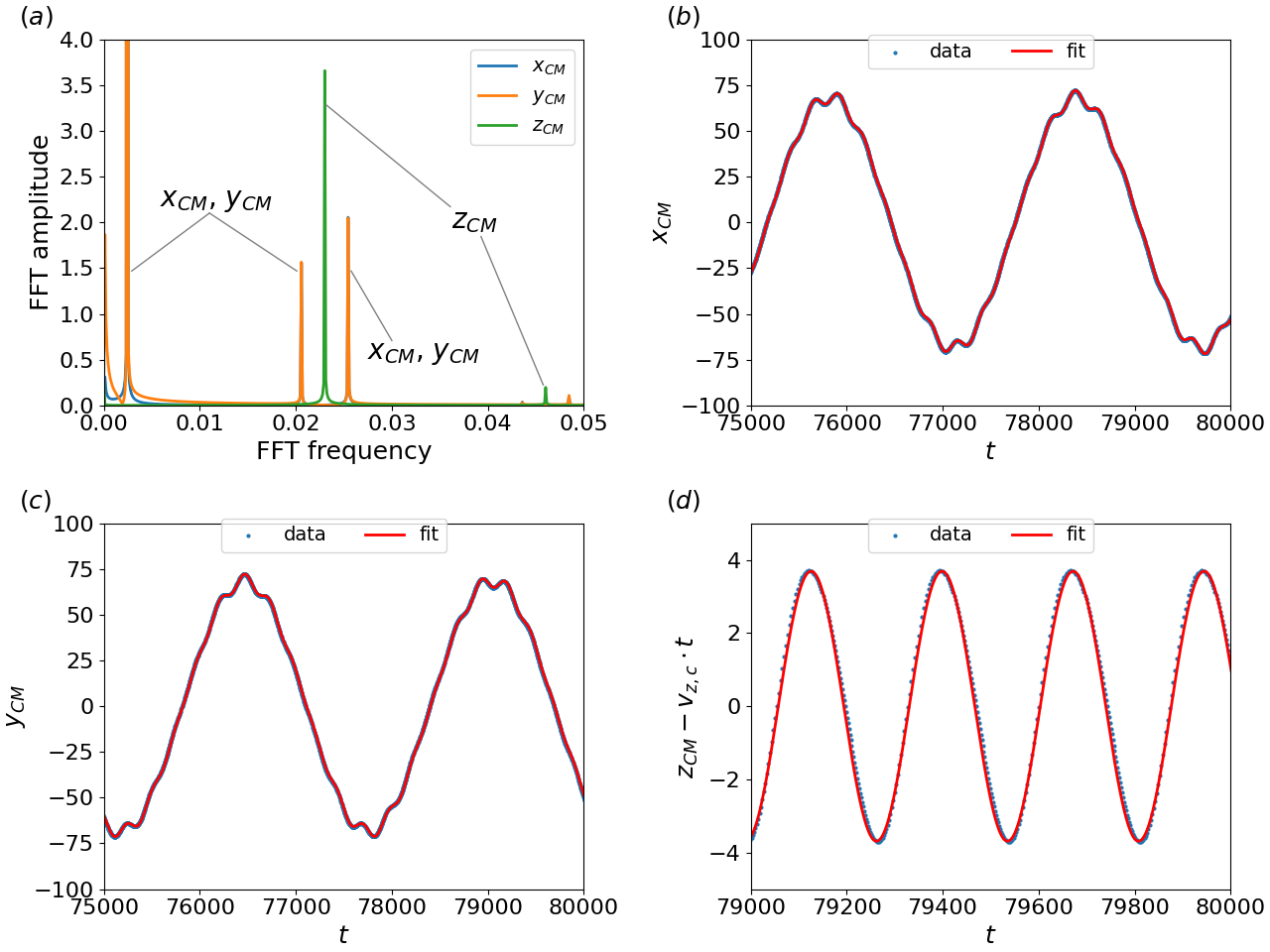}
    \vspace{-0.2cm}
    \caption{
    The centre-of-mass movement for the GRTT mode with $B=16362$. 
    (\textit{a})~The FFT of $x_{CM}$, $y_{CM}$ (overlapped) and $z_{CM}$. 
    (\textit{b})-(\textit{c})~The fit of quasi-periodic $x_{CM}$ and $y_{CM}$ was performed with three dominant FFT frequencies, and the amplitude of the first frequency, equal to $68.60$, is truncated in the plot. 
    (\textit{d})~The fit of periodic $z_{CM}-v_{z,c}t$ was performed with one FFT frequency only.
    See \eqref{eq:grtt_cm} for the fitting functions. 
    }
    \label{fig:fit_grtt_cm}
\end{figure}

The results of the FFT analysis of the data and the fitting expressions for the GRTT mode with $B=16362$ are shown in figures~\ref{fig:fit_grtt_angles}-\ref{fig:fit_grtt_cm}. 
Even though the amplitudes of the frequencies $\omega_r-\omega_g$ and $\omega_r+\omega_g$ are more than 30 times smaller than the amplitude of $\omega_g$, they need to be taken into account to demonstrate the quasi-periodicity of $x_{CM}(t)$ and $y_{CM}(t)$. 
The parameters are 
$\theta_0 = 47 \degree$, 
$\omega_g = 2.43 \cdot 10^{-3}$, %
$\omega_r = 2.30 \cdot 10^{-2}$, %
and 
$v_{z,c} = 1.489$.

\clearpage

\section{The critical value $B_{crit}$ for the transition between the rocking and GRTT modes}
\label{app:GRTT_Bcrit}
\begin{figure}
    \centering
    \includegraphics[width=0.45\textwidth]{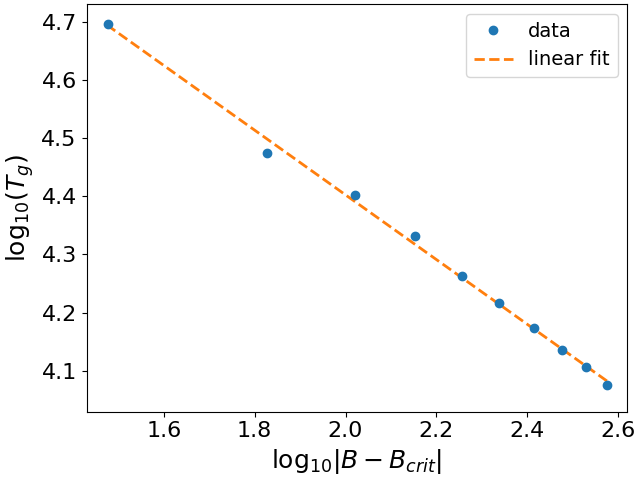}
    \vspace{-0.2cm}
    \caption{
    Log-log plot of the gyration period $T_{g}$ in the GRTT mode as a function of $|B-B_{crit}|$ confirms a power law dependence \eqref{Tg} with the critical value $B_{crit}=14218$. 
    The range of elasto-gravitation number $B$ chosen for the fit is $14248 \leq B \leq 14594$.
    }
    \label{fig:fit_grtt_Bcrit}
\end{figure}

Figure~\ref{fig:fit_grtt_Bcrit} demonstrates that in the GRTT mode, when $B$ 
decreases to a critical value $B_{crit}$, the gyration period $T_g$ diverges as $T_g \sim |B-B_{crit}|^{-\chi}$, in agreement with \eqref{Tg}.
The fitting procedure was used to determine the critical value of $B_{crit}$. 
The range of $B$ for which this procedure was performed was limited to the closest data points near the expected $B_{crit}$: $14248 \leq B \leq 14594$ 
(see figure~\ref{fig:period_grtt_cm_gyration}(a)). 
The best fit, which yielded the smallest residual, resulted in $\chi = -0.5560$ and $B_{crit} = 14218$. 
For larger values, the GRTT mode is observed, and for smaller values, the rocking mode is present.

\newpage
\section{Examples of stationary and non-stationary planar configurations of a sedimenting elastic loop}
\label{no-circle}

\begin{figure}
    \centering    
    \includegraphics[width=0.45\textwidth]{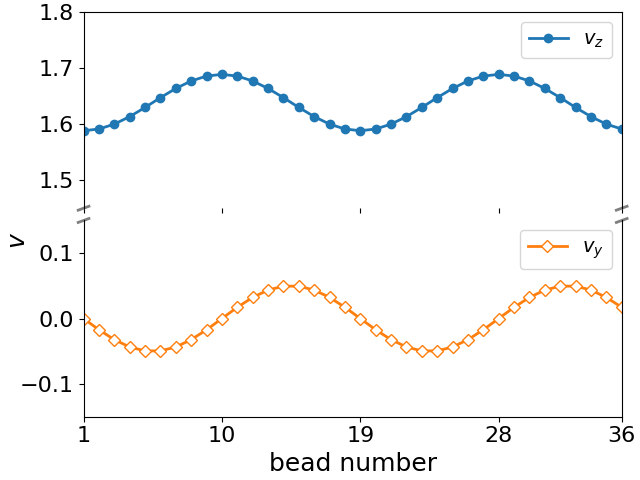}
    \vspace{-0.2cm}
    \caption{A circular vertical loop in the elastic equilibrium is not stationary.  
    Vertical ($v_z$, dots) and horizontal ($v_y$, diamonds) components of the $i$-th  bead velocity $\bm{v}_{i}^{\parallel}$ vary with the bead number $i$. 
    }
    \label{fig:vertical_circle}
\end{figure}

First, we demonstrate that a vertical circle made of $N$ beads is not a stationary configuration, even if it is in elastic equilibrium. 
For $N=36$, the beads 19 and 1 are at the top and bottom, respectively, and the line of centres of the beads 10 and 28 is horizontal.
The numerically evaluated $i$-th bead velocities are shown in Figure \ref{fig:vertical_circle}. 
The vertical component of the 1-st and 19-th bead velocity is smaller than for the 10-th and 28-th beads, by approximately 6\%. The horizontal component of velocity tends to move apart the upper beads and the lower beads closer to each other. 

Therefore, the circular vertical configuration of the elastic loop is not stationary. This conclusion can be supported by an analytical calculation within the point-particle model. %
We will show that the sedimentation velocity of the top bead is smaller than the sedimentation velocity of the right side bead, ${v}_{19}%
< {v}_{10}%
$
. %
The contribution to ${v}_{19}%
$ from four beads: $19\!+\!k$, $19\!-\!k$, and their horizontal  reflections, $1\!+\!k$, $37\!-\!k$, scaling as $v_k^v=\cos\phi_k + 1/\cos\phi_k + \sin\phi_k + 1/\sin\phi_k$  is smaller than the contribution to ${v}_{10}$ from the analogous rotated configuration of four beads $10\!+\!k$, $10\!-\!k$, and their vertical  reflections, $28\!+\!k$, $28\!-\!k$, scaling as $v_k^h=\sin^2\phi_k/\cos\phi_k + 1/\cos\phi_k + \cos^2\phi_k/\sin\phi_k + 1/\sin\phi_k$,  where $k=1,...,8$ and $\phi_k= \pi(18-k)/36 > \pi/4$. 
The contribution to ${v}_{10}$ from the beads 1 and 19 is the same as the contribution ${v}_{19}$ from the beads 10 and 28. 
The contribution to ${v}_{19}$ from the bead 1, scaling as $1$, is larger than the contribution to ${v}_{10}$ from the bead 28, scaling as $1/2$, but it is easy to check that this difference is smaller than the difference $v_1^h-v_1^v$.   
Therefore, within the point-particle model, ${v}_{19}%
< {v}_{10}%
$. 

The reasoning presented above is generalized straightforwardly for an inclined circle other than horizontal -- the vertical gravitational force needs to be replaced by its projection parallel to the circle's plane. 
Therefore, a circle other than horizontal is not a stationary configuration. 

\begin{figure}
    \centering    
    \includegraphics[width=0.91\textwidth]{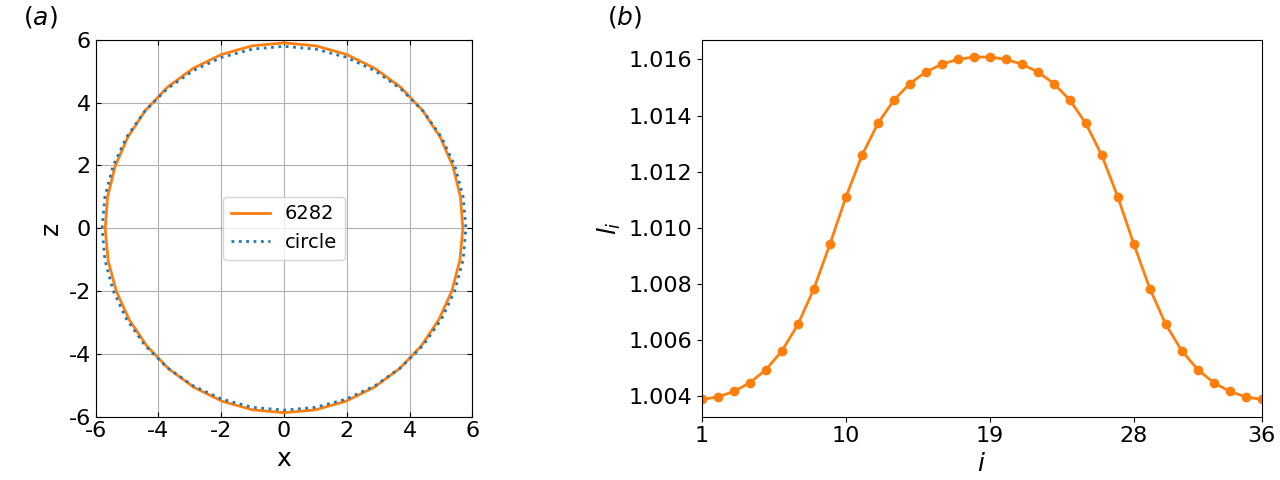}
    \vspace{-0.2cm}
    \caption{
    The planar vertical stationary shape of the elastic loop in the attracting vertical mode for $N$=36 \& $B$=6282 (solid line in (\textit{a})) is shown to differ from a circle (dotted line in (\textit{a})). 
    The distance between the consecutive beads $l_i$ depends on the bead position  -- it is the smallest at the bottom, and the largest at the top, as visible in (\textit{b}), with 
     $l_i=|\bm{r}_{i+1}-\bm{r}_{i}|$ for $i=1,...,N-1$, and $l_N=|\bm{r}_{1}-\bm{r}_{N}|$.
    The distance in the elastic equilibrium equals to 1.01. 
    }
    \label{fig:vertical_shape}
\end{figure}

\begin{figure}
    \centering
    \includegraphics[width=0.45\textwidth]{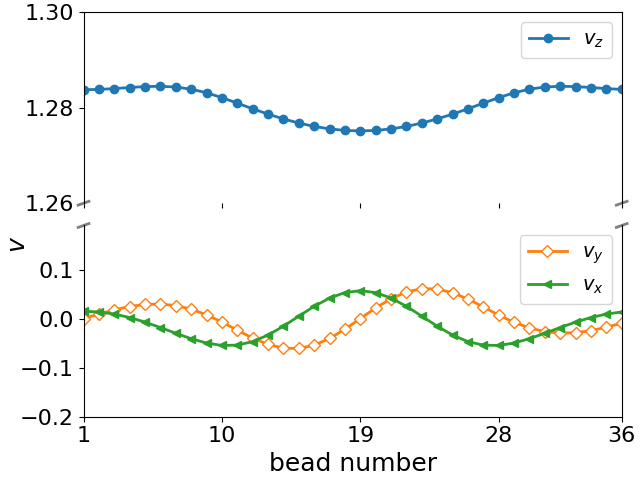}
    \vspace{-0.2cm}
    \caption{Velocities of the beads for the planar shape from figure~\ref{fig:vertical_shape} placed horizontally ($z \rightarrow y$). This configuration is not stationary. 
    }
    \label{fig:horizontal_shape}
\end{figure}

Then we show that the stationary planar vertical configuration of the attracting vertical mode is non-circular, and it is not stationary if placed horizontally. 
The non-circular, vertical shape of the attracting vertical mode, visible in figure~\ref{fig:vertical_shape}, was also reported by \cite{gruziel-slomka_stokesian_2019}. 
This shape is not stationary if placed horizontally -- velocities of all the beads are not the same, as shown in figure~\ref{fig:horizontal_shape}. 
This conclusion can be easily generalized for an inclined loop other than vertical if the gravitational forces are replaced by their projections perpendicular to the loop's plane. 
Therefore, the planar configuration observed in the attracting vertical mode is not stationary after tilting by an angle $\theta \ne \pi/2$. 

Finally, we argue that the horizontal circular stationary configuration of the loop in the elastic equilibrium is unstable.  Indeed, the elastic loop with $B=6283$, initially horizontal with the shape shown in figure \ref{fig:vertical_shape},  evolves towards a horizontal circle, but then it moves out of the horizontal plane.

\newpage
\section{Comparisons with the dynamics of sedimenting elastic loops in \cite{gruziel-slomka_stokesian_2019}}
\label{cGS}

The results obtained here for the elastic loops made of $N=36$ beads of diameter $d$, with their centres at the elastic equilibrium separated by $1.01d$,  will be now compared to \cite{gruziel-slomka_stokesian_2019}. 
The same loops are modelled there as an elastic ring made of 60 overlapping beads of the same diameter $d$, but with their centres at the elastic equilibrium separated by $0.6d$, which provides approximately the same geometrical aspect ratio. %
The elasto-gravitation number $B$, used in this work and defined in \eqref{Bdef}, relates in the following way to the stiffness parameter $\tilde{A}$ used in \cite{gruziel-slomka_stokesian_2019} with the correction in \cite{gruziel-slomka_correction_2022},
\begin{equation}
    B \approx 1.5 \cdot 10^4 /{\tilde{A}}.
\end{equation}
The above comparison is based on assuming \cite{bukowicki_different_2018} that the loop from \cite{gruziel-slomka_stokesian_2019, gruziel-slomka_correction_2022} has the same bending energy 
\begin{equation}
   U^{B} = \sum_{j=1}^{60} \frac{A^{*}}{1.2d} \beta'^2_{j}, 
   \label{UBbis}
\end{equation}
as in \eqref{UB} the loop  with $N=36$ considered here. 
In \eqref{UBbis}, $A^*$ from \cite{gruziel-slomka_stokesian_2019} is the same bending stiffness as $A$ used here.

In \cite{gruziel-slomka_stokesian_2019}, hydrodynamic interactions between the beads were described by the Rotne-Prager translational-translational mobility matrix~\cite{Rotne1969,Yamakawa1970,Zuk2014,zuk_intrinsic_2017,zuk2018}. This approach does not include the lubrication forces caused by the relative motion of close bead surfaces (in contrast to our method). To compensate, in \cite{gruziel-slomka_stokesian_2019}, the consecutive beads overlap, and small short-distance repelling forces are added. 
This approach allows for monitoring the motion of very elastic loops, such as the bent figure eight or toroidal attracting modes. 
Such compact modes cannot be studied by our present numerical approach. 
Another advantage of the Rotne-Prager method is that the simple analytical form of the mobility coefficients allows for fast computations.
The benefit of the multipole expansion corrected for lubrication (used here) is a controlled high precision. 

 The attracting dynamical modes (vertical, tilted, frozen rotating, swinging, flapping) observed in the Rotne-Prager approximation \cite{gruziel-slomka_stokesian_2019} have also been found here. The transition between vertical and tilted modes takes place at approximately the same value of $B$. The coexistence of different modes, found in \cite{gruziel-slomka_stokesian_2019}, is confirmed here. However, in the present work, two new modes and some differences in the phase diagram are reported. Moreover, here %
 all the modes are described quantitatively, including the characteristic time scales, velocities, and transitions between the modes. %

\bibliographystyle{unsrt}  

\bibliography{references.bib}

\end{document}